\documentclass[superscriptaddress,twocolumn,amsmath,amssymb,pra,floatfix,noshowpacs]{revtex4-1}
\usepackage{graphicx}
\usepackage{dcolumn}
\usepackage{bm}

\usepackage{color}
\usepackage{ulem}
\usepackage{tikz}
\usepackage{rotating}
\usepackage{amsmath}
\usepackage{amssymb}

\begin{document}

\title{Unconventional supersymmetric quantum mechanics in spin systems}
\author{Amin Naseri}
\email{amin.naseri@csu.edu.cn}
\affiliation{School of Physics and Electronics, Central South University, Changsha, P. R. China 410083}

\author{Yutao Hu}
\affiliation{School of Physics and Electronics, Central South University, Changsha, P. R. China 410083}

\author{Wenchen Luo}
\email{luo.wenchen@csu.edu.cn}
\affiliation{School of Physics and Electronics, Central South University, Changsha, P. R. China 410083}

\begin{abstract}
It is shown that the eigenproblem of any $2\times 2$ matrix Hamiltonian with discrete eigenvalues is involved with a supersymmetric quantum mechanics. The energy dependence of the superalgebra marks the disparity between the deduced supersymmetry and the standard supersymmetric quantum mechanics. The components of an eigenspinor are superpartners\textemdash up to a $SU(2)$ transformation\textemdash which allows to derive two reduced eigenproblems diagonalizing the Hamiltonian in the spin subspace. As a result, each component carries all information encoded in the eigenspinor. We also discuss the generalization of the formalism to a system of a single spin-$\frac{p}{2}$ coupled with external fields. The unconventional supersymmetry can be regarded as an extension of the Fulton-Gouterman transformation, which can be established for a two-level system coupled with multi oscillators displaying a mirror symmetry. The transformation is exploited recently to solve Rabi-type models. Correspondingly, we illustrate how the supersymmetric formalism can solve spin-boson models with no need to appeal a symmetry of the model. Furthermore, a pattern of entanglement between the components of an eigenstate of a many-spin system can be unveiled by exploiting the supersymmetric quantum mechanics associated with single spins which also recasts the eigenstate as a matrix product state. Examples of many-spin models are presented and solved by utilizing the formalism.\end{abstract}
\maketitle
\section{Introduction}
Quantum entanglement is turned out to be a crucial property in targeting relevant corners of the Hilbert space of a many-body system \cite{Vidal01,ORUS,Verstraete,Eisert,Vidal02} rather than merely being a puzzle in foundations of quantum mechanics \cite{QE}. Entanglement is perceived as a correlation between constituents of a composite system. Specifically, an entangled state of a bipartite system is a quantum state which cannot be described as a product of states of subsystems \cite{QE}. A two-level system, i.e. a single spin-$\frac{1}{2}$, interacting with an external field provides a concrete example in this situation. The setting is among the simplest quantum systems which models varieties of physical phenomena including atom-field interactions and quantum optics \cite{Rabi,Scully}, mesoscopic and molecular physics \cite{Wagner,Thanopulos}, etc.. The eigenstates of such models are expected to describe an entanglement between the spin and the external field if no good spin quantum number exists.

Our focus here is on models which can be partitioned into two parts and their eigenproblem (EP) is involved with a $2 \times 2$ matrix Hamiltonian. The aim of this paper is to show that there exist redundancies in degrees of freedom (DOF) required to construct the eigenspinor of the system if the state is entangled. The entanglement also allows to represent a two-component eigenspinor of the system by means of the only one of the components. We generalize the formalism to a single spin-$\frac{p}{2}$ for $p> 1$ coupled with external fields. In these setups also the entanglement reduces DOF required to describe the eigenstate of the system. One step further, we show a pattern of entanglement can be revealed in an eigenstate of a many-spin system by taking into account the redundancies associated with single spins. This allows to reconstruct the eigenstate as a matrix product state (MPS) \cite{Verstraete,ORUS}.

An essential ingredient of our formalism is connected with the supersymmetric quantum mechanics (SUSY QM) \cite{Bagchi,Junker,Cooper}. Supersymmetry (SUSY) is conjectured in relativistic quantum field theory for unification of fundamental forces of nature. The prediction of superpartners of elementary particles, whose spins differ by a half-integer, has not been observed experimentally. This implies SUSY has to be spontaneously broken. SUSY was initially introduced into non-relativistic quantum mechanics to study the non-perturbative breaking of SUSY \cite{WITTEN} and later became a framework on its own. In addition to proposals for its realization in condensed matter and cold atom systems \cite{ezawa,Yu,Tomka:2015aa,Hsieh,Bradlyn,Lahrz}, SUSY QM is adopted in classifying analytically solvable potentials in quantum mechanics in a close relation with the factorization method \cite{Dong}.

Let us briefly review SUSY QM and mention the associated aspects which are of interest to our study. SUSY QM can be constructed by two supercharge operators $Q_{\pm}=F_{\pm} \sigma_{\pm}$ and a Hamiltonian $\mathrm{H}=Q_{+}+Q_{-}$, where $2 \sigma_{\pm}=\sigma_{x}\pm i \sigma_{y}$ with $\sigma_{x,y,z}$ being Pauli matrices and $F^{}_{\pm}=F^{\dag}_{\mp}$ are some external fields. The supercharges satisfy $Q^{2}_{\pm}=0$. The EP of the Hamiltonian is $\mathrm{H} |\boldsymbol{\phi} \rangle = E |\boldsymbol{\phi} \rangle$ where
 \begin{equation}
| \boldsymbol{\phi} \rangle= \begin{pmatrix}
     | \phi_{+} \rangle   \\
      | \phi_{-} \rangle
\end{pmatrix}
 \end{equation}
is an eigenspinor and $E$ is the associated eigenenergy. The square of the Hamiltonian $\mathrm{H}^{2}$ has a diagonal form and together with the supercharges close the superalgebra
 \begin{equation}
 	\mathrm{H}^{2}=\{Q_{+},Q_{-}\}, \quad \left[Q_{\pm},\mathrm{H}^{2} \right]=0,
 \end{equation}
where $\{,\}$ is an anticommutator and $\left[,\right]$ is a commutator. $\mathrm{H}^{2}$ has a good supersymmetry if its ground states $|\boldsymbol{\phi}_{0} \rangle$ with $E=0$ are vacuums of all supercharges $Q_{\pm}|\boldsymbol{\phi}_{0} \rangle =0$. $|\boldsymbol{\phi}_{0} \rangle$ can be written as a product of an eigenspinor of $\sigma_{z}$ and the vacuum of either $F_{+}$ or $F_{-}$, and hence, is a disentangled state. The components of an eigenspinor with $E\neq0$ are superpartners $F_{\pm} |\phi_{\mp} \rangle=E |\phi_{\pm} \rangle$ and by knowing one of the components, the other can be found. If $F_{\pm}$ are non-hermitian operators, then
\begin{equation}
	|\boldsymbol{\phi} \rangle=
	\begin{pmatrix}
     \frac{1}{E}F_{+}| \phi_{-} \rangle   \\
     | \phi_{-} \rangle 
	\end{pmatrix}=
	\begin{pmatrix}
     | \phi_{+} \rangle   \\
     \frac{1}{E} F_{-}| \phi_{+} \rangle   
	\end{pmatrix}
\end{equation}
is clearly an entangled state since it cannot be written as a product of an eigenspinor of $\sigma_{z}$ and a state associated with the fields $F_{\pm}$. The intertwining relationship between components leads to two EPs $F_{\pm} F_{\mp}| \phi_{\pm} \rangle=E^{2}| \phi_{\pm} \rangle$. Solving either of the EPs solves the EP of $H$. We generalize these considerations to a generic $2\times 2$ matrix Hamiltonian and show these bipartite systems are endowed with a SUSY QM. The associated superalgebra is defined in terms of energy-dependent operators, and hence, is dubbed unconventional supersymmetric quantum mechanics (USUSY QM) in order to distinguish it from the standard SUSY QM. The existence of supercharges of the USUSY QM makes it possible to solve the EP of the full system via solving one of the EPs associated with its subsystems. That is a subsystem of the system contains all the information of the system due to the correlation between DOF.

The formalism presented herein can also be regarded as an extension to the Fulton-Gouterman (FG) transformation \cite{FulGout}. A two-level system interacting with multi oscillators can be diagonalized by the FG transformation in the spin subspace if the Hamiltonian possesses a mirror symmetry. Particularly in the context of spin-boson models, different generalizations of the Rabi model are solved by exploiting the FG transformation inspired by the approach introduced in the solution of the Rabi model \cite{Braak}. The USUSY formalism relaxes the symmetry constraint of the FG transformation.

The FG transformation is generalized to a multi-level system interacting with a multi-oscillatory system also restricted to Hamiltonians with a discrete symmetry \cite{Wagner_1984}. The generalization of the FG transformation can also be covered by the USUSY formalism irrespective of the symmetry of the model. This means a Hamiltonian describing interactions between a single spin-$\frac{p}{2}$, i.e. a $(p+1)$-level system, and external fields can be diagonalized in the spin subspace due to redundancies associated with the USUSY QM independent of symmetries of the Hamiltonian. The EP can be then reduced to $p+1$ decoupled EPs. Moreover, solving one of $p+1$ EPs solves the full EP. This connects also our work to the parasupersymmetric quantum mechanics \cite{Cooper,Bagchi} where supercharge operators are generalized to parasupersymmteric charges in a diagonal Hamiltonian associated with a single spin-$\frac{p}{2}$. However, our approach here is to reduce DOF of aforementioned systems by means of constraints associated with USUSY QM. 

The structure of the paper and a brief summary of our results are as follows. In Sec.~\ref{sec_USUSY}, we introduce USUSY QM at a generic ground. In Sec.~\ref{subsec_half}, we present the USUSY QM pertaining to the EP of a generic $2\times 2$ matrix Hamiltonian. In Sec.~\ref{subsec_phalf}, we demonstrate that the eigenproblem of a Hamiltonian describing interactions between some fields and a single spin-$\frac{p}{2}$ for $p>1$ is also endowed with a USUSY QM. In Sec.~\ref{sec_manyS}, we devise USUSY QM associated with single spins to reduce DOF of a generic interacting many-spin system.

Examples are presented in Sec.~\ref{sec_ex} and Sec.~\ref{sec_MSE} to illustrate how USUSY QM can be exploited in dealing with the EP of spin systems. We set the \textit{direct exact diagonalization} (ED) as a benchmark. By direct ED we mean the diagonalization of a Hamiltonian represented in a basis constructed out of disentangled states. A single spin-$\frac{1}{2}$ in a uniform magnetic field is the simplest Hamiltonian endowed with an USUSY QM which is considered in Sec.~\ref{susec_UBShalf}. In Sec.~\ref{susec_JC}, we solve the Jaynes-Cummings (JC) model \cite{JC,JC50}. The treatment of Rabi-type models \cite{Xie} are presented in Sec.~\ref{susec_gRabi}. In Sec.~\ref{susec_2S}, we start to analyze systems including more than one spin by solving a model of interacting spins and mention how the formalism can be engaged iteratively if the system is extended by adding more spins. In Sec.~\ref{susec_TC} and Sec.~\ref{susec_gD}, we analyze the eigensolutions of Tavis-Cummings (TC) model \cite{TC} in a MPS representation and solve generalized Dicke models \cite{Dicke,Kirton}, respectively, which describe many spins interacting with a single boson mode.

\section{Unconventional Supersymmetric Quantum Mechanics}\label{sec_USUSY}
In this section, we construct the USUSY QM for the EP of a $2\times 2$ matrix Hamiltonian. Next, the scheme is generalized to a model describing interactions between a single spin-$\frac{p}{2}$ with $p>1$ and external fields. Finally, we discuss an interacting system of many spin-$\frac{1}{2}$ and reconstruct its eigenstates by means of USUSY QM associated with the single spins.   
\subsection{Single Spin-$\frac{1}{2}$}\label{subsec_half}
We start from a generic $2\times 2$ matrix Hamiltonian and the associated EP 
\begin{equation}
	H
	=
	\begin{pmatrix}
     H_{+} &F_{+}    \\
    F_{-} &H_{-}  
\end{pmatrix},  \quad 
H |\Psi \rangle=E |\Psi \rangle,
\label{eq_GSHam}
\end{equation}
where $H_{\pm}=H_{\pm}^{\dag}$, $F_{+}=F_{-}^{\dag}$ and the eigenstate is a two-component spinor $\langle \Psi |=(\langle \psi_{+} | \,\, \langle \psi_{-} |)$. $H_{\pm}$ are restricted to those with discrete eigenvalues $E^{(n)}_{\pm}$, collected in sets $\mathcal{E}_{\pm}$. If $E \notin \mathcal{E}_{\pm}$, the components are certainly connected
\begin{equation}
 | \psi_{\pm} \rangle=G_{\pm} F_{\pm} | \psi_{\mp} \rangle, \quad G_{\pm} =(E-H_{\pm})^{-1},
\end{equation}
which allow to construct the eigenstate from one of the components
\begin{equation}
	|\Psi \rangle =
\begin{pmatrix}
     G_{+} F_{+} | \psi_{-} \rangle   \\
     | \psi_{-} \rangle 
\end{pmatrix}
=\begin{pmatrix}
      | \psi_{+} \rangle   \\
     G_{-} F_{-} | \psi_{+} \rangle 
\end{pmatrix}.
\label{eq_SDis}
\end{equation}
The EP of $H$ is therefore decoupled into two EPs
\begin{eqnarray}
	h_{\pm}
	 | \psi_{\pm} \rangle
	  =
	  E  | \psi_{\pm} \rangle,
	 \quad
	h_{\pm} = H_{\pm}
	+F_{\pm}
	G_{\mp}
	F_{\mp},
	\label{eq_decoupled01}
\end{eqnarray}
which show components belong to subsystems governed by $h_{\pm}$ and either carries all information encoded in $|\Psi \rangle$. $h_{\pm}$ do not represent physical systems as they have eigenstates which are associated with more than one eigenenergy. The induced potentials $F_{\pm} G_{\mp}	F_{\mp}$ compensate the decoupling of the components and can turn $h_{\pm}$ into a non-local differential equation in problems having a position space representation
\begin{equation}
	\langle \mathbf{r} | F_{\pm} G_{\mp}	F_{\mp} |\mathbf{r}^{\prime} \rangle
	\neq \langle \mathbf{r} | F_{\pm} G_{\mp}	F_{\mp} |\mathbf{r} \rangle \delta(\mathbf{r}-\mathbf{r}^{\prime}),
\end{equation}
where $\delta(\mathbf{r}-\mathbf{r}^{\prime})$ is the Dirac delta function. The EPs and $h_{\pm}$ in Eq.~\eqref{eq_decoupled01} are addressed as reduced EPs and reduced Hamiltonians, respectively, in the following. Equivalently, the original EP can be presented with a diagonal matrix
\begin{equation}
	\begin{pmatrix}
     h_{+} &0    \\
    0 &h_{-}  
\end{pmatrix}  \begin{pmatrix}
     | \psi_{+} \rangle   \\
     | \psi_{-} \rangle 
\end{pmatrix}=E \begin{pmatrix}
     | \psi_{+} \rangle   \\
     | \psi_{-} \rangle 
\end{pmatrix}.
\end{equation}
The fact that $| \psi_{\pm} \rangle$ are degenerate under the action of the hermitian operators $h_{\pm}$ implies a SUSY QM. To make it explicit, we define two supercharge operators
\begin{equation}
	\mathcal{Q}_{\pm}=G_{\pm}F_{\pm}\sigma_{\pm},
	\label{eq_such01}
\end{equation}
which fulfil $\mathcal{Q}_{\pm}^{2}=0$ independent of the nature of external fields coupled with the spin. The EP of $H$ can now be recast as $\mathcal{H} |\Psi \rangle=|\Psi \rangle$ by a traceless matrix $\mathcal{H}=\mathcal{Q}_{+}+\mathcal{Q}_{-}$ which acts like an idempotent operator $\mathcal{H}^{2} |\Psi \rangle=\mathcal{H} |\Psi \rangle $. $\mathcal{Q}_{\pm}$ and $\mathcal{H}^{2}$ satisfy a superalgebra
\begin{eqnarray}
	\{\mathcal{Q}_{+},\mathcal{Q}_{-} \}=\mathcal{H}^{2},   \quad \left[\mathcal{Q}_{\pm},\mathcal{H}^{2} \right]=0.
	\label{eq_SUSYAL}
\end{eqnarray}
The SUSY QM is characterized by energy-dependent operators, and therefore, it is dubbed USUSY QM in order to distinguish it from the standard SUSY QM. The existence of the USUSY QM, like a symmetry, reduces DOF of the problem as only one of the reduced EPs is needed to be dealt.

The energy-dependent operators can lead to a breakdown of the USUSY QM, which contrasts the difference between the standard SUSY QM and USUSY QM. This can happen for a given eigenstate with $E \in \mathcal{E}_{+}$ or $E \in \mathcal{E}_{-}$. However, a unitary transformation of the Hamiltonian
\begin{equation}
\tilde{H}= U^{\dag} H U =\begin{pmatrix}
      \tilde{H}_{+}&  \tilde{F}_{+} \\
      \tilde{F}_{-}   &\tilde{H}_{-}
\end{pmatrix},
\quad
U \in  SU(2)
\label{eq_SU2}
\end{equation}
leading also to
\begin{equation}
|\tilde{\Psi} \rangle= U^{\dag} 
|\Psi \rangle
=
\begin{pmatrix}
     |\tilde{\psi}_{+} \rangle\\
     |\tilde{\psi}_{-} \rangle
\end{pmatrix}, \quad
\tilde{G}_{\pm} =(E-\tilde{H}_{\pm})^{-1},
\end{equation}
induces new diagonal terms which can have different sets of eigenenergies $\tilde{\mathcal{E}}_{\pm}\neq {\mathcal{E}}_{\pm}$. As far as $\tilde{H}_{\pm}$ have discrete eigenenergies, the singularity can be removed as $U$ changes the eigenenergies of the diagonal terms continuously. Consequently, the singular points are isolated and the USUSY QM can be realized for any eigenstate of $H$ if $U$ warrants $E \notin \tilde{\mathcal{E}}_{\pm}$.

In retrospect, assume we solve $h_{+}$ in order to find eigensolutions of $H$. Those eigenenergies of $H$ which are in $\mathcal{E}_{-}$ might be missing. That is $E \in  \mathcal{E}_{-}$ is a necessary condition for having a missing solution of $H$ though it is not a sufficient condition. We note also that the excluded solutions are those states with no USUSY QM. If
\begin{equation}
H |\Psi \rangle = E^{(m)}_{-} |\Psi \rangle,
\quad  \langle m|\psi_{-} \rangle \neq 0,
\label{eq_exc}
\end{equation}
where $H_{-} |m \rangle= E^{(m)}_{-}  |m \rangle$, are satisfied, then solutions of $h_{+}$ exclude $E^{(m)}_{-}$. In fact, Eq.~\eqref{eq_exc} sets a constraint on parameters of the model if $E^{(m)}_{-}$ is an eigenenergy of $H$. A class of excluded solutions, which can be readily figured out, are those states satisfying
\begin{equation}
F_{+} |0 \rangle^{}_{F}=0, \quad H_{-} |0 \rangle^{}_{F}=E |0 \rangle^{}_{F},
\end{equation}
that is the vacuum of $F_{+}$ is an eigenstate of $H_{-}$. The eigenstate can be chosen as $| \Psi \rangle=  | - \rangle| 0 \rangle^{}_{F}   $ where $\sigma_{z}| \pm \rangle=\pm | \pm \rangle $. The spin and the field are disentangled in these states clearly and the states are an eigenspinor of $\sigma_{z}$. However, not all the excluded solutions are disentangled states and not all disentangled eigensolutions are excluded. Excluded entangled eigenstates require a fine-tuning of parameters of the model and need to be looked for in models with no parametric analytical solutions.

In view of a perturbation theory, if off-diagonal terms of $H$ are considered as a perturbation to the diagonal terms, the excluded solutions of $H$ are non-perturbative solutions as the sum of all non-zero order corrections has to vanish. In some problems, the states without a USUSY QM can be shown to be among states with isolated exact solution, e.g. the Rabi model in the single-mode spin-boson representation \cite{Emary}. Another approach to check whether there exists any missing solution is to compare the solutions of different $\tilde{h}_{+}$ corresponding to some different unitary transformations of $H$ in Eq.~\eqref{eq_SU2} which is practical if solutions within a finite range of energy are of interest. If $h_{-}$ is chosen to be solved instead, the above consideration can be applied as well by a swap in the notation $+ \leftrightarrow  -$. 

The existence of USUSY QM is a strong condition for reducibility of an EP and it can be relegated to the existence of either $Q_{\pm}$ as it supports the existence of either $h_{\mp}$, correspondingly. Specifically, one needs to transform $H$ to find a representation in which one of the intertwining relationships $| \tilde{\psi}_{\pm} \rangle=\tilde{G}_{\pm} \tilde{F}_{\pm} | \tilde{\psi}_{\mp} \rangle$ can be constructed for those eigenstates of $H$ lying within a given energy range of interest. In other words, the EP can be reduced if the components are maximally correlated at least through one way, e.g. $| \psi_{+} \rangle=G_{+} F_{+} | \psi_{-} \rangle$, then knowing $| \psi_{-} \rangle$ uniquely establishes $| \psi_{+} \rangle$. If one of the intertwining relationships exists, we address it as a reducible EP and if none of them exist, it is regarded as an irreducible EP. The components are considered to be superpartners if both intertwining relationships exist.

\subsection{Single Spin-$\frac{p}{2}$}\label{subsec_phalf}
We generalize the formalism to deal with a spin-$\frac{p}{2}$ for $p>1$. A Hamiltonian describing interactions between a single spin-$\frac{p}{2}$ and external fields has the following form
\begin{eqnarray}
	H_{ {  \frac{p}{2}}}
	=H_{0}+\Delta S_{z}+F_{+} S_{+}+F_{-} S_{-},
	\label{eq_SPhalf}
\end{eqnarray}
where $H_{0}$ does not include spin operators. The fields satisfy $F_{+}=F^{\dag}_{-}$ and $\Delta=\Delta^{\dag}$. $S_{z}$ and $S_{\pm}$ are spin-$\frac{p}{2}$ generators of $SU(2)$ algebra \cite{Bagchi,Cooper} 
\begin{eqnarray}
	( S_{+} )_{jk}&=&\sqrt{j(p-j+1)} \delta_{j+1,k},\\
	S_{z}&=&diag\left(p,p-2,\cdots, -p+2,-p\right),
\end{eqnarray}
for $1\leqslant j ,k\leqslant p+1$, which satisfy
\begin{eqnarray}
	[S_{+},S_{-}]= S_{z}, \quad
	[S_{z},S_{\pm}]= \pm 2S_{\pm}.
\end{eqnarray}
The EP is $H_{\frac{p}{2}} |\Psi \rangle=E|\Psi \rangle$ where $E$ is the associated eigenenergy and $|\Psi \rangle$ has $p+1$ components. The scheme of making the Hamiltonian traceless by inverting the diagonal terms in the EP $\mathcal{H}_{ {  \frac{p}{2}}} |\Psi \rangle=|\Psi \rangle$ and then squaring it $\mathcal{H}^{2}_{ {  \frac{p}{2}}} |\Psi \rangle=|\Psi \rangle$ results as well here in decoupling (almost) half of the components of the eigenspinor from the rest. This is because $\mathcal{H}_{ {  \frac{p}{2}}}$ can be written as a sum of two supercharges
\begin{equation}
	\mathcal{Q}^{(\frac{p}{2})}_{\pm}=\mathrm{I}_{\pm} \mathcal{K} \left(S_{+}+S_{-} \right), 
	\quad \text{with} \quad
	\left(\mathcal{Q}^{(\frac{p}{2})}_{\pm} \right)^{2}=0,
	\label{eq_such02}
\end{equation}
where $\mathrm{I}_{\pm}$ and $ \mathcal{K}$ are diagonal matrices with entries 
\begin{equation}
(\mathrm{I}_{+})_{jj}=\delta_{j,2l+1}, \quad (\mathrm{I}_{-})_{jj}=\delta_{j,2l}, 
\end{equation}
given $l$ is an integer number, and
\begin{equation}
	(\mathcal{K})_{jj}=\left[E-H_{0}-\Delta(p-2j+2)\right]^{-1}.
\end{equation}
We note that the supercharges in Eq.~\eqref{eq_such02} reduce to the spin-$\frac{1}{2}$ supercharges in Eq.~\eqref{eq_such01} for $p=1$. The scheme can be repeated over and over on the components which are not decoupled in order to achieve ultimately $p+1$ decoupled EPs (an example is worked out in App. \ref{app_S1}). Once again, a unitary transformation can turn an irreducible eigenstate to a reducible one by lifting the degeneracy between $H_{\frac{p}{2}}$ and diagonal terms.
\subsection{Many spin-$\frac{1}{2}$}\label{sec_manyS}
It is possible to explore how the intertwining relationship between components of single spins can be exploited to reduce DOF in a many-spin system. We illustrate it for the case of a system of many spin-$\frac{1}{2}$ and show a MPS \cite{Verstraete,ORUS} emerges automatically after removing the redundancies. Furthermore, a pattern of entanglement also appears through reconstructing the many-spin eigenstate by means of the intertwining relationship of single spins. A Hamiltonian describing interactions between many spins and external fields reads
\begin{equation}
H^{\scriptscriptstyle {  (N)}}=H_{0}+\sum_{{ k}=1}^{N}\left( \Delta^{ (k)} \sigma_{z}^{ (k)}+F^{(k)}_{+} \sigma_{+}^{ (k)}+F^{(k)}_{-} \sigma_{-}^{ (k)} \right),
	\label{eq_HN01}
\end{equation}
where here also $H_{0}$ does not include $\sigma_{x,y,z}^{ (k)}$ for $k=1, \cdots, N$. The nature of fields coupled to the spins does not affect our results. For instance, a one-dimensional chain of $N+1$ spins can also be represented in a similar form as $H^{\scriptscriptstyle {  (N)}}$ if the fields are replaced by
\begin{equation}
F^{(k)}_{+}=\sigma_{-}^{(k+1)}=(F^{(k)}_{-})^{\dag} \quad \text{and} \quad \Delta^{ (k)}=\sigma_{z}^{(k+1)}.
\end{equation}
Spin models on a $D$-dimensional lattice can also be represented similarly given $k$ is replaced by a proper vector. In order to exercise USUSY QM of a single spin, $H^{\scriptscriptstyle {  (N)}}$ can be represented, for instance, in eigenspinors of $\sigma_{z}^{ (1)}$
\begin{equation}
	 H^{ { \scriptscriptstyle (N)}}=
	\begin{pmatrix}
      H^{\scriptscriptstyle{  (N-1)}}_{+}&F_{+}^{(1)}   \\
      F_{-}^{(1)}   &  H^{ \scriptscriptstyle{  (N-1)}}_{-}
\end{pmatrix},
	\label{eq_NHam}
\end{equation}
where $H^{\scriptscriptstyle{  (N-1)}}_{\pm}=H^{\scriptscriptstyle{  (N-1)}} \pm \Delta^{(1)}$ and
\begin{equation}
H^{\scriptscriptstyle{  (N-1)}}=H_{0}+\sum_{{ k}=2}^{N}\left( \Delta^{\scriptscriptstyle (k)} \sigma_{z}^{\scriptscriptstyle (k)}+F^{\scriptscriptstyle (k)}_{+} \sigma_{+}^{ \scriptscriptstyle(k)}+F^{\scriptscriptstyle (k)}_{-} \sigma_{-}^{ \scriptscriptstyle (k)} \right) .
\nonumber
\end{equation}
An eigenspinor of $H^{ \scriptscriptstyle{  (N)}}$ can be grouped into two components $	\langle \Psi^{ \scriptscriptstyle{  (N)}} | =
	(
  \langle \psi^{ \scriptscriptstyle{  (N-1)}}_{{  +}} | \,\,
     \langle \psi^{ \scriptscriptstyle{  (N-1)}}_{{  -}} | )$. The same procedure adopted for a single spin-$\frac{1}{2}$ decouples $|\psi^{ \scriptscriptstyle{  (N-1)}}_{{  \pm}} \rangle  $ in the corresponding EP associated with the eigenenergy $E$ and renders two reduced Hamiltonians
\begin{equation}
	h_{\pm}^{\scriptscriptstyle{  (N-1)}}
	=
	H^{\scriptscriptstyle{  (N-1)}}_{\pm}+F_{\pm}^{\scriptscriptstyle (1)} \left(E-H^{\scriptscriptstyle{  (N-1)}}_{\mp} \right)^{-1}F_{\mp}^{\scriptscriptstyle (1)}
\end{equation}
if $|\Psi^{ \scriptscriptstyle{  (N)}} \rangle$ is reducible. That is by solving an $(N-1)$-spin subsystem, the system of $N$ spins is solved. It is worthwhile to mention that in the case of $N=2$, the same procedure adopted for a single spin-$\frac{p}{2}$ for $p>1$ works as well here in decoupling four components of the eigenspinor and achieving four reduced EPs, but it fails for $N>2$. 

\begin{figure}[tb]
\begin{centering}
\begin{tikzpicture}
\begin{scope}[xshift=0.00\textwidth,yshift=0.23\textwidth]
\node[anchor=south west,inner sep=0](image) at (0,0){
\includegraphics[width=0.45\textwidth]{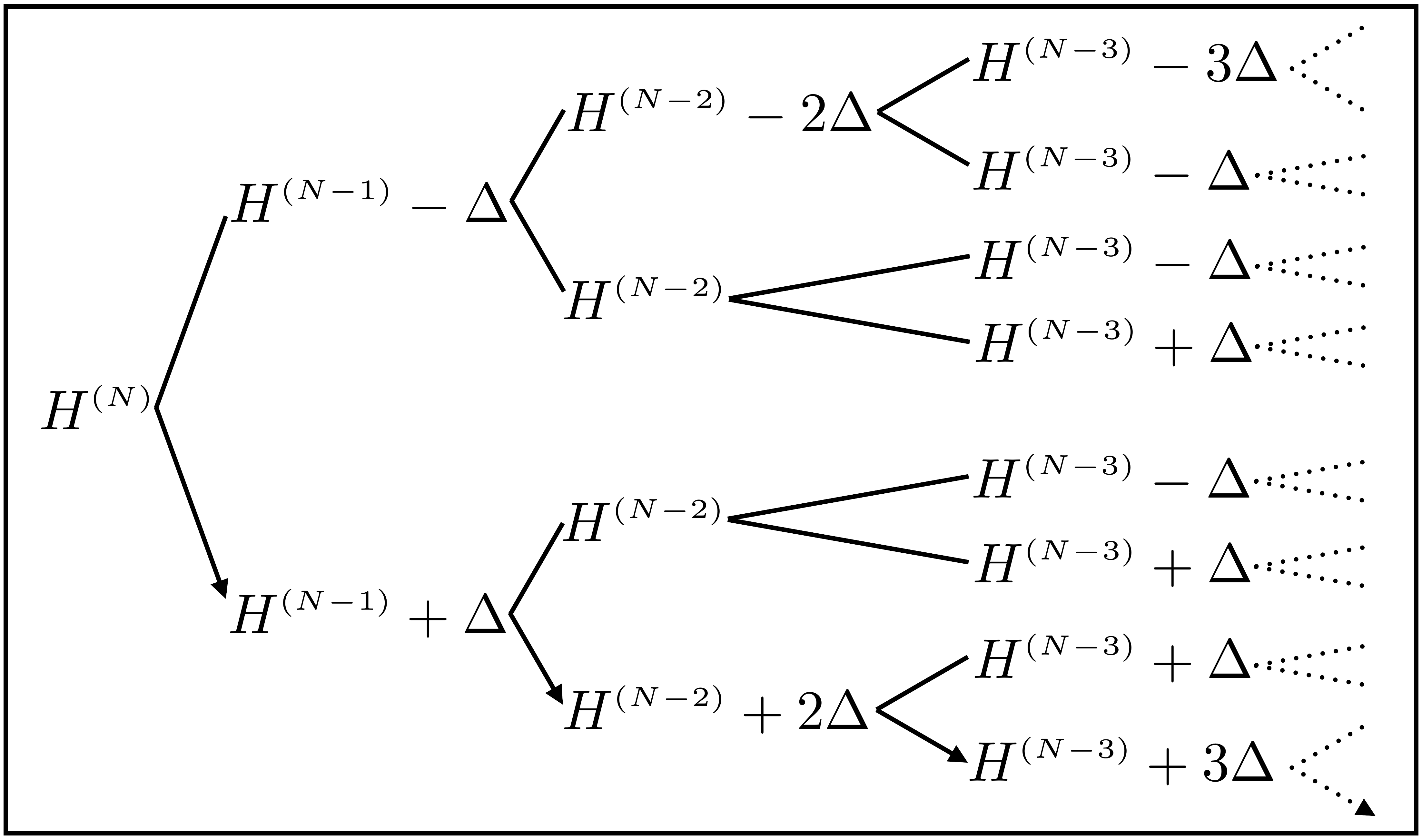}};
\end{scope}
\end{tikzpicture}
\par\end{centering}
\caption{The possible paths for decomposing the eigenstates of $H^{\scriptscriptstyle {  (N)}}$ where $F^{(k)}_{\pm}=F_{\pm}$ and $\Delta^{(k)}=\Delta$ are assumed for simplicity. By representing $H^{\scriptscriptstyle {  (N)}}$ in eigenspinors of $\sigma_{z}^{(1)}$, eigenstates of $H^{\scriptscriptstyle (N-1)}+ \Delta$, for instance, can be used to represent an eigenstate of $H^{\scriptscriptstyle {  (N)}}$ through the relation in Eq.~\eqref{eq_SDis}. The same procedure can be carried out by representing $H^{\scriptscriptstyle (N-1)}+ \Delta$ in the eigenspinors of $\sigma_{z}^{(2)}$ and expanding its eigenstates in terms of eigenstates of $H^{\scriptscriptstyle (N-2)}+ 2\Delta$. The recursive procedure can be followed until $|\Psi^{ \scriptscriptstyle{  (N)}} \rangle$ finds an expansion in terms of eigenstate of $H^{\scriptscriptstyle (0)}+ N\Delta$. The arrows show the path which is adopted for decomposing $|\Psi^{ \scriptscriptstyle{  (N)}} \rangle$ into eigenstates of subsystems.}
\label{SConf}
\end{figure}

Using the relation in Eq.~\eqref{eq_SDis}, $|\Psi^{ \scriptscriptstyle{  (N)}} \rangle$ can be constructed out of one of the components, for example $|\psi^{ \scriptscriptstyle{  (N-1)}}_{-} \rangle  $. $|\psi^{ \scriptscriptstyle{  (N-1)}}_{-} \rangle  $ can be expanded in terms of eigenstates of $H^{\scriptscriptstyle{  (N-1)}}_{+}$ or $H^{\scriptscriptstyle{  (N-1)}}_{-}$ since both Hamiltonians contain $N-1$ spins and provide a proper basis. Suppose we choose the eigenstates of $H^{\scriptscriptstyle{  (N-1)}}_{+}$ as a basis. By representing $H^{\scriptscriptstyle{  (N-1)}}_{+}$ in the eigenspinor of $\sigma_{z}^{(2)}$, once again the relation in Eq.~\eqref{eq_SDis} can be used to reconstruct the reducible eigenstates of $H^{\scriptscriptstyle{  (N-1)}}_{+}$ in terms of one of the components. The irreducible eigenstates of $H^{\scriptscriptstyle{  (N-1)}}_{+}$ with a non-zero projection on $|\psi^{ \scriptscriptstyle{  (N-1)}}_{-} \rangle  $ are stuck and cannot be connected to smaller spin subsystems. Similar to the single-spin case, a unitary transformation can alter the set of irreducible states and remove them from the representation. The same procedure can be repeated $N$ times to exploit USUSY QM of all spins which attributes a fractal structure to $|\Psi^{ \scriptscriptstyle{  (N)}} \rangle$ as presented in the following. We assume the spins are coupled with an identical field $F^{\scriptscriptstyle (k)}_{\pm}=F_{\pm}$ and $\Delta^{\scriptscriptstyle (k)}=\Delta$ as this simplification does not change the pattern of entanglement which we are interested to reveal. According to the hierarchical structure of $H^{\scriptscriptstyle{  (N)}}$, there is no unique path for decomposing $|\Psi^{ \scriptscriptstyle{  (N)}} \rangle$ in terms of eigenstates of the subsystems. We pursue the path indicated in Fig.~\ref{SConf} by arrows. 

If no irreducible state comes up in the course of repeating the scheme down to the level where USUSY QM of all spins are exploited, $|\psi^{ \scriptscriptstyle{  (N-1)}}_{-} \rangle  $ can be represented formally as
\begin{equation}
	|\psi^{ \scriptscriptstyle (N-1)}_{  -} \rangle =
	\sum_{\boldsymbol{j}}
	\Omega_{j^{}_{  N-1}} 
	\prod^{  N-1}_{\ell =1} 
	\Omega_{  j^{}_{  N-\ell-1}}^{  j^{}_{  N-\ell}}
	|\varsigma^{}_{\boldsymbol{j}} \rangle,
	\label{eq_NSS}
\end{equation}
where $\boldsymbol{j}=\{ j^{}_{ 0}, \cdots, j^{}_{  N-1}\}$ and $j^{}_{N-\ell}$ is a collective quantum number labelling the eigenstate $|j^{}_{N-\ell}   \rangle$ and the associated eigenenergy $E^{}_{j^{}_{N-\ell}}$ of $H^{\scriptscriptstyle{  (N-\ell)}} + \ell\Delta$. All quantum numbers designated by $j^{}_{N-\ell}$ for $1 \leqslant \ell \leqslant  N$ are summed up in Eq.~\eqref{eq_NSS}. The coefficient of the expansion are grouped into tensors, i.e. one vector and $N-1$ matrices, with elements
\begin{eqnarray}
\Omega_{j^{}_{ \scriptscriptstyle N-1}}&=&\langle j^{}_{ \scriptscriptstyle  N-1}|\psi^{  \scriptscriptstyle (N-1)}_{  -}  \rangle, 
	\label{eq_MPC01}
	\\
	\Omega_{  j^{}_{\scriptscriptstyle N- \ell}}^{  j^{}_{N- \ell+1}}
	&=&
	\langle j^{}_{ \scriptscriptstyle N- \ell} | \Theta_{  j^{}_{\scriptscriptstyle N-\ell+1}}^{\scriptscriptstyle{  (N-\ell)}} \rangle,
	\label{eq_MPC02}
\end{eqnarray}
where
\begin{equation}
| j^{}_{\scriptscriptstyle N-\ell+1} \rangle=
\begin{pmatrix}
| \Phi_{  j^{}_{N-\ell+1}}^{\scriptscriptstyle{  (N-\ell)}} \rangle   \\
| \Theta_{  j^{}_{N-\ell+1}}^{\scriptscriptstyle{  (N-\ell)}} \rangle
\end{pmatrix}.
\end{equation}
The eigenstates are explicitly reconstructed by the relation in Eq.~\eqref{eq_SDis} as
\begin{equation}
	|j^{}_{\scriptscriptstyle N-\ell+1} \rangle=
	g^{}_{  j^{}_{  N-\ell+1}} 
	\begin{pmatrix}
	|\Theta_{  j^{}_{N-\ell+1}}^{\scriptscriptstyle{  (N-\ell)}}  \rangle  \\
	|\Theta_{  j^{}_{N-\ell+1}}^{\scriptscriptstyle{  (N-\ell)}} \rangle
	\end{pmatrix}
\end{equation}
with
\begin{eqnarray}
g^{}_{  j^{}_{  N-\ell+1}}&=& diag(\mathcal{G}^{\scriptscriptstyle{  (N-\ell)}}_{ j^{}_{  N-\ell+1}}F^{}_{+},1), 
\\
\mathcal{G}^{\scriptscriptstyle{  (N-\ell)}}_{ j^{}_{  N-\ell+1}}
	&=&
	 (E_{ j^{}_{  N-\ell+1}}-H^{\scriptscriptstyle{  (N-\ell)}}-\ell \Delta)^{-1}. 
\end{eqnarray}
In Eq.~\eqref{eq_NSS}, $|\varsigma^{}_{\boldsymbol{j}} \rangle$ is defined as 
\begin{equation}
	|\varsigma^{}_{\boldsymbol{j}} \rangle
	=
	\prod^{  N-1}_{\ell =1} I^{}_{   \ell-1} 
	\otimes 
	g^{}_{  j^{}_{  N-\ell}} 
	|j_{  0}^{  } \rangle	|\mathbf{ 1}\rangle,
	\label{eq_NSS02}
\end{equation}
where $ I_{  \ell}$ is a $2^{\ell} \times 2^{\ell}$ identity matrix, $|j_{  0}^{} \rangle$ is an eigenstate of $H^{ \scriptscriptstyle{  (0)}}_{+}=H_{{  0}}+N \Delta $, and
\begin{equation}
	|\mathbf{ 1} \rangle=\bigotimes_{k=2}^{N} \left(|+\rangle_{k} +|-\rangle_{k} \right), \,\, \text{where} \,\, \sigma_{z}^{(k)}|\pm\rangle_{k}=\pm|\pm\rangle_{k}.
	\nonumber
\end{equation}
The expansion in Eq.~\eqref{eq_NSS} can be understood in this way that subsystems of the $N$-spin system provide intrinsically \textit{a correlated orthonormal basis} to represent the $N$-spin eigenstate. The coefficients of the expansion in Eq.~\eqref{eq_MPC02} are specified by two indices which indicates they are entries of some matrices $\boldsymbol{\Omega}^{\scriptscriptstyle (N-\ell)}$. Therefore in other words, an $N$-spin eigenstate is represented as a MPS by removing the redundancies associated with the single spins. No approximation is imposed in the derivation of the representation. So the problem after removing the redundancies is identical to the original problem modulo local transformations to remove irreducible eigenstates. It is shown in \cite{Vidal04} that every state can be represented in a MPS representation. Our result is consistent with this statement in the case of eigenstates (opposed to an arbitrary state). However, the MPS emerges here through exploiting the correlations provided by the Hamiltonian rather than applying Schmidt decomposition \cite{Vidal04}.

If the dimension of the matrices $\boldsymbol{\Omega}^{\scriptscriptstyle (N-\ell)}$ is assumed to be $\lesssim \chi$ for instance by a truncation, the number of coefficients to fix the $N$-spin eigenstate is $\propto \chi^{2} \times N$ which grows linearly rather than exponentially with number of spins. The correlated basis should be contrasted with a basis constructed out of disentangled states with a dimension of $2^{ \scriptscriptstyle N}\times M$ to diagonalize $H^{ \scriptscriptstyle{  (N)}}$, where $M$ is the number of states required to represent the fields coupled to spins. If the correlated basis is used to diagonalize the reduced Hamiltonian $h_{  -}^{\scriptscriptstyle{  (N-1)}}$, a considerably smaller (truncated) basis is required to reproduce the results of a direct ED. This is plausible since correlations between DOF are partially accumulated in the eigenbases of the subsystems. Furthermore, the expansion in Eq.~\eqref{eq_NSS} reveals the pattern of correlations between components of an eigenspinor of $H^{ { \scriptscriptstyle (N)}}$, and can be used to construct a relevant ansatz. A similar representation can be established for a system of many spin-$\frac{p}{2}$ due to the USUSY QM associated with the single spins.
\section{Single-spin examples}\label{sec_ex}
In this section, we solve some single-spin models by their reduced EPs in order to illustrate the formalism. First, we solve the EP of a spin in a uniform magnetic field. Next, JC model and Rabi-type models are studied which are involved with boson creation (annihilation) operator $a^{\dag}$ ($a$) with $[a,a^{\dag}]=1$. The unit of energy is set equal to unity $\hbar \omega=1$ in all examples herein.
\subsection{A single spin-$\frac{1}{2}$ in a uniform magnetic field}\label{susec_UBShalf}
We begin by considering an elementary example of a single spin-$\frac{1}{2}$ interacting with a constant magnetic field $\mathbf{B}_{0}=(B_{1},B_{2},B_{3})$ which is described by the EP
\begin{equation}
\mathbf{B}_{0} \cdot \boldsymbol{\sigma} |\boldsymbol{\sigma} \rangle=E|\boldsymbol{\sigma} \rangle,
\end{equation}
where $\boldsymbol{\sigma} =(\sigma_{x},\sigma_{y},\sigma_{z})$ and $\langle \boldsymbol{\sigma} |=( c^{\ast}_{+}\,\, c^{\ast}_{-} )$ with $c_{\pm} \in\mathbb{C}$. As far as $B_{1,2}\neq 0$, either of the reduced EPs in Eq.~\eqref{eq_decoupled01} gives both eigenvalues
\begin{eqnarray}
	&&\left( \pm B_{3}+\frac{B_{1}^{2}+B_{2}^{2}}{E \mp B_{3}}\right)c_{\pm}=E c_{\pm} \nonumber \\&& \Rightarrow E_{\pm}=\pm\sqrt{B_{1}^{2}+B_{2}^{2}+B_{3}^{2}}.
\end{eqnarray}
Non-normalized eigenspinors then can be constructed by Eq.~\eqref{eq_SDis} as
\begin{equation}
	|\boldsymbol{\sigma}^{}_{\pm} \rangle
	=
	\begin{pmatrix}
      \frac{B_{1}-iB_{2}}{ E_{\pm}-B_{3}}    \\
      1  
\end{pmatrix},
\end{equation}
which are orthogonal $\langle \boldsymbol{\sigma}^{}_{-}|\boldsymbol{\sigma}^{}_{+} \rangle=0$.

\subsection{Jaynes-Cummings model}\label{susec_JC}
The reduced EPs in Eq.~\eqref{eq_decoupled01} can be algebraically solved if $H_{\pm}$ and $F_{\pm}$ close under commutation relations. The celebrated JC model \cite{JC,JC50} is a suitable example to demonstrate it. The JC model describes interaction between a single spin and a single boson mode. The Hamiltonian in Eq.~\eqref{eq_GSHam} becomes the JC Hamiltonian $H_{JC}$ by choosing $H_{\pm}=a^{\dag} a \pm \Delta$ and $F_{+}=\alpha a$ with $\Delta$ and $\alpha$ being real numbers. The conserved operator of the JC Hamiltonian $C=a^{\dag} a+\frac{1}{2}\sigma_{z}$ induces $[h_{\pm},a^{\dag}a]=0$ which indicates eigenstates of $h_{\pm}$ are the eigenstates of the number operator $a^{\dag} a | n\rangle  = n | n\rangle $. The common eigenenergies of $h_{\pm}$ can be found by solving
\begin{equation}
 (E-n-1+\Delta)(E-n-\Delta)=\alpha^{2}(n+1), \,\, n \in \mathbb{N},
 \label{eq_JCSec}
 \end{equation}
which gives two solutions for a given $n$. Parity quantum numbers $\eta=\pm$ are required to label the eigenenergies
\begin{equation}
E^{(n,\eta)}=n+\frac{1}{2}+\eta \sqrt{(\Delta-\frac{1}{2})^{2}+\alpha^{2}(n+1)}
\end{equation}
and the corresponding non-normalized eigenstate
\begin{equation}
|\Psi^{(n,\eta)} \rangle
=	\begin{pmatrix}
     |\psi_{+}^{(n,\eta)}    \rangle\\
      |\psi_{-}^{(n,\eta)} \rangle
\end{pmatrix}
=	
\begin{pmatrix}
     |n    \rangle\\
      \frac{\alpha}{E^{(n,\eta)}-n-1+\Delta}|n+1\rangle
\end{pmatrix}.
\end{equation}
The full eigenstate can be constructed either by an eigenstate of $h_{+}$ or the one of $h_{-}$ since the components are superpartners. The only eigenstate with no USUSY QM is the disentangled state $\langle \Psi^{d} |=(0 \,\,\,  \langle 0 | )$ with $E^{d}=-\Delta$ which is the vacuum of $\mathcal{Q}_{+}$, but $\mathcal{Q}_{-}$ does not exist for this state, and the state has no parity number. Therefore, all the eigensolutions of JC model can be found by solving $h_{-}$. Although $H_{JC}$ and $H_{+}$ can be tuned to have common eigenenergies, no eigenenergy of $H_{JC}$ is missed by solving $h_{-}$ for $\alpha \neq 0$. This is because
\begin{equation}
H_{JC} |\Psi \rangle=(m+\Delta)|\Psi \rangle, \quad \langle m |\psi_{+} \rangle\neq 0
\end{equation}
for a given $m$ cannot be satisfied. It is notable that the normalized spin texture
\begin{equation}
\bar{ \rho}^{(n,\eta)}_{x}(r)=\frac{\left( \Psi^{(n,\eta)}(r)\right)^{\dag} \sigma_{x} \Psi^{(n,\eta)}(r)}{|\Psi^{(n,\eta)}(r)|^{2}}
\end{equation}
with $
 \Psi^{(n,\eta)}(r)=\langle r|\Psi^{(n,\eta)} \rangle
$ has a topological index \cite{PRB,Weinberg} defined by
\begin{equation}
\lim_{r \rightarrow \infty} \bar{ \rho}^{(n,\eta)}_{x}(-|r|) \bar{ \rho}^{(n,\eta)}_{x}(|r|)\rightarrow-1,
\end{equation}
which is absent in $\Psi^{d}(r)$. This implies the non-local aspect of the entanglement in the JC model.
\subsection{Rabi-type models}\label{susec_gRabi}
If $H_{\pm}$ and $F_{\pm}$ in Eq.~\eqref{eq_GSHam} do not close an algebra, the reduced EP can be expanded in a given basis and the eigenenergies are the roots of the determinant of the matrix. If the Hilbert space is not finite, a truncated basis can be used to achieve low-energy eigensolutions which is equivalent to the approach of ED routine. This is particularly proper Rabi-type models.

\begin{figure}[tb]
\begin{centering}
\begin{tikzpicture}
\begin{scope}[xshift=0.00\textwidth,yshift=0.23\textwidth]
\node[anchor=south west,inner sep=0](image) at (0,0){
\includegraphics[width=0.48\textwidth]{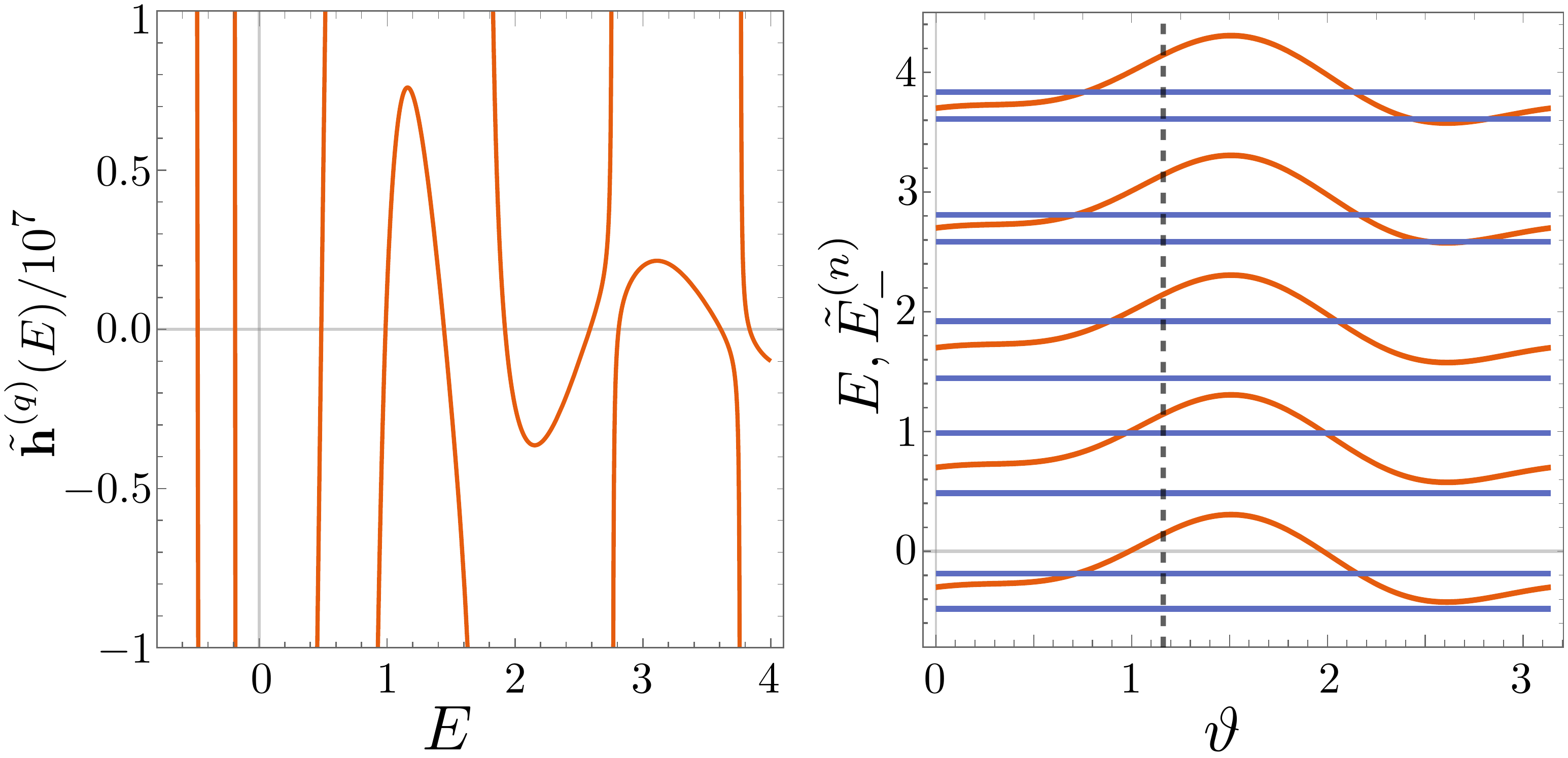}};
\end{scope}
\end{tikzpicture}
\par\end{centering}
\caption{(color online) Generalized Rabi model's results for $\alpha =0.7$, $\beta=0.3$, $\Delta=0.3$ and $\gamma=0.2 e^{i \pi/3}$. Left: $\tilde{\mathbf{h}}^{(q)}(E)$ versus $E$ is plotted for $q=12$ and the transformation is fixed by $\tan{(\vartheta/2)}=\sqrt{\beta/\alpha}$. Right: The horizontal lines are eigenenergies of the generalized Rabi model found by a direct ED. The curvy lines are $\tilde{E}_{-}^{(n)}$ for $n=0,1, \cdots , 9$ which are a function of $\vartheta$. At the crossing points, USUSY QM might not exist for the corresponding eigenstate of the generalized Rabi model. The vertical dashed line is the angle associated with $\alpha =0.7$, $\beta=0.3$.}
\label{GR1}
\end{figure}

We consider a generalization of the Rabi model \cite{Xie} in the presence of a homogeneous in-plane magnetic field, specified by $H_{\pm}=a^{\dag} a \pm \Delta $ and $F_{+}=\alpha a+\beta a^{\dag}+\gamma$ in Eq.~\eqref{eq_GSHam}, in order to illustrate solving a spin-boson model without a closed algebra. Without loss of generality, we suppose $\alpha,\beta \geqslant 0$ while $\gamma \in \mathbb{C}$. For $\gamma=0$, the reduced Hamiltonians posses a parity symmetry as $[h_{\pm},e^{i a^{\dag} a \pi}]=0$ and the intertwining relationship implies two components have different parities $\langle \psi_{+}|\psi_{-} \rangle=0$. If the reduced Hamiltonian has a tridiagonal matrix representation, an analytical progress can be made as the determinant of the matrix representation can be achieved by a three-term recurrence equation \cite{horn}. The reduced Hamiltonians of the generalized Rabi model are tridiagonal if at least one of $\alpha$, $\beta$, $\gamma$ is zero. Otherwise, a transformation
\begin{equation}
U\equiv U(\vartheta) =e^{i \sigma_{y} \frac{\vartheta}{2} }\,\, \text{with} \,\, \tan{ \frac{\vartheta}{2}}=\left( \frac{\beta}{\alpha}\right)^{\pm\frac{1}{2}}
\end{equation}
makes the representation of either $\tilde{h}_{\pm}$ tridiagonal, see App.~\ref{app_TMRGR}. We proceed with $\tilde{h}_{+}$ and assume $|\tilde{n} \rangle$ are the eigenstates of $\tilde{H}_{-}$ associated with the eigenenergy $\tilde{E}_{-}^{(n)} $ after rotating $H$ with $\tan{(\vartheta/2)}=\sqrt{\beta/\alpha}$. Making the matrix out of elements
\begin{equation}
\tilde{f}^{(m,n)}=\langle \tilde{m} |( \tilde{h}_{+}-E)|\tilde{n} \rangle, 
\end{equation}
the $j$-th continuant $\tilde{\mathbf{h}}^{(j)}(E)$ obeys a recurrence relation \cite{horn}
\begin{equation}
	\tilde{\mathbf{h}}^{(j+1)}(E)=\tilde{f}^{(j,j)} \tilde{\mathbf{h}}^{(j)}(E)-|\tilde{f}^{(j,j-1)}|^{2}
	 \tilde{\mathbf{h}}^{(j-1)}(E)
	 \label{eq_recu}
\end{equation}
with initial values $\tilde{\mathbf{h}}^{(-1)}(E)=0$ and $\tilde{\mathbf{h}}^{(0)}(E)=1$. Given the desired convergence is achieved by keeping $q+1$ states, the eigenvalues of $\tilde{h}_{+}$ are the roots of the characteristic polynomial $\tilde{\mathbf{h}}^{(q)}(E)=0$, see Fig.~\ref{GR1}.

As discussed earlier, if $E=\tilde{E}_{-}^{(m)}$ and $\langle \tilde{m} |\tilde{\psi}_{-} \rangle \neq 0$, where $\langle \tilde{\Psi}|=(\langle \tilde{\psi}_{+}| \,\, \langle \tilde{\psi}_{-}|)$ is an eigenstate of $\tilde{H}$, then the eigensolution would be excluded from the solutions of $\tilde{\mathbf{h}}^{(q)}(E)=0$. We consider the possibility of having excluded solutions from different aspects. First, we check for which $\vartheta$ the full Hamiltonian $H$ and $\tilde{H}_{-}$ can have a common eigenvalue within a given range of energy while parameters $\alpha,\beta, \gamma$ and $\Delta$ are fixed numerically. Low-lying eigenenergies of $\tilde{H}_{-}$ together with the eigenenergies of $H$ are plotted versus $\vartheta$ in Fig.~\ref{GR1} which shows there is no degeneracy between $H$ and $\tilde{H}_{-}$, and hence, no missing solution within the range of energy, for a transformation with $\vartheta=2 \arctan{\sqrt{\beta/\alpha}}$ with $\alpha=0.7$ and $\beta=0.3$. It is notable that at least for the ground state, minimizing $\tilde{E}_{-}^{(0)}$ with respect to $\vartheta$ provides a better approximation in comparison with a zero-order perturbation theory on $H$, i.e. $\vartheta =0$, while taking the off-diagonal terms as a perturbation to the diagonal terms.

\begin{figure}[tb]
\begin{centering}
\begin{tikzpicture}
\begin{scope}[xshift=0.00\textwidth,yshift=0.23\textwidth]
\node[anchor=south west,inner sep=0](image) at (0,0){
\includegraphics[width=0.38\textwidth]{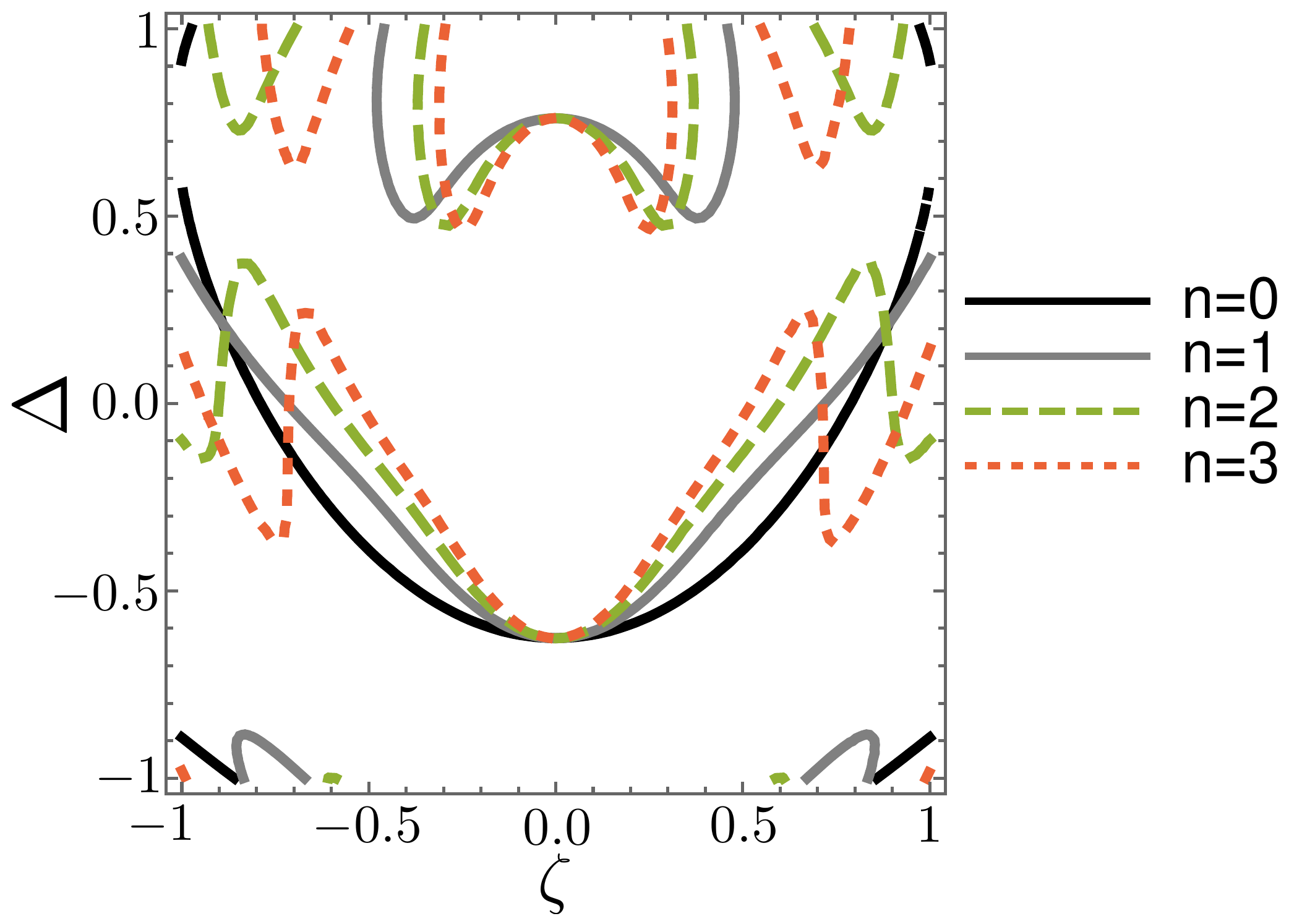}};
\end{scope}
\end{tikzpicture}
\par\end{centering}
\caption{(color online) Generalized Rabi model's results. The relation between $\Delta$ and $\zeta$ (defined by $\alpha =\zeta \cosh{\tau}$ and $\beta=\zeta \sinh{\tau}$ with $\tanh{\tau}=3/7$) is given if $\tilde{H}|\tilde{\Psi} \rangle =\tilde{E}_{-}^{(n)} |\tilde{\Psi} \rangle $ is satisfied for $n=0,1,2,3$, and $\gamma=0.2 e^{i \pi/3}$ within a direct ED treatment.}
\label{GR1b}
\end{figure}

On the other hand, it is possible to find out the interrelationship between parameters of the model by diagonalizing $\tilde{H}|\tilde{\Psi} \rangle =\tilde{E}_{-}^{(n)} |\tilde{\Psi} \rangle $, which provides the necessary condition for a missing solution. The results for some $n$ are depicted in Fig.~\ref{GR1b}. Finally, exact isolated solutions \cite{Judd,Emary,Xie} also can be constructed for the eigenstates without USUSY QM. In this class of solutions the expansion of $|\tilde{\psi}_{\pm} \rangle$ in terms of the eigenstates of either $\tilde{H}_{\pm}$ have a finite number of terms and the eigenenergy is fixed to either $E=\tilde{E}^{(n)}_{\pm}$, correspondingly, for a given $n$, see App.~\ref{app_IES}. The exact solution needs to be sought self-consistently by deriving a set of equations which constrain the parameters of the model.

Braak derived the spectrum of the Rabi model
\begin{equation}
	H_{R}=
	\begin{pmatrix}
      a^{\dag}a +\Delta&  g(a+a^{\dag}) \\
      g(a+a^{\dag})  &a^{\dag}a -\Delta 
\end{pmatrix}
\end{equation}
by virtue of the Bargmann-Fock space \cite{Braak} (some models solved by exploiting the same formalism are reviewed in \cite{Xie}). The regular eigenenergies (contrasted with isolated exact solutions including Judd points \cite{Judd}) are zeros of a transcendental function i.e. the so-called $G$-function. Each summand in the definition of the $G$-function is associated with a sequence which needs to be evaluated separately through a three-term recurrence relation which is involved with the tridiagonality of the Rabi model. Finding roots of the $G$-function also requires a truncation similar to the ED treatment applied here. $\mathbb{Z}_{2}$ symmetry of the Rabi model is invoked through the FG transformation \cite{FulGout,Wagner_1984,Moroz} to reduce DOF in Braak's approach. 

The same results can be reproduced by employing the USUSY QM of the Rabi model as is presented briefly in the following. The reduced Hamiltonian of the Rabi Hamiltonian
\begin{equation}
	h_{R}=
	a^{\dag}a +\Delta 
	+
	(a+a^{\dag})\frac{g^{2}}{E-a^{\dag}a +\Delta}(a+a^{\dag})
\end{equation}
in the associated reduced EP $h_{R} |\psi_{R} \rangle=E|\psi_{R} \rangle  $ needs to be solved and then the eigenstate of $H_{R}$ would be
\begin{equation}
	|\Psi_{R} \rangle=
	\begin{pmatrix}
      |\psi_{R} \rangle \\
      g\left(E-a^{\dag}a +\Delta\right)^{-1} (a+a^{\dag})|\psi_{R} \rangle
\end{pmatrix}.
\end{equation}
In order to exploit the $\mathbb{Z}_{2}$ symmetry of $H_{R}$, the reduced EP needs to be expanded either in the even-parity basis $|n_{e} \rangle =|2n \rangle$ or the odd-parity basis $|n_{o} \rangle =|2n+1 \rangle$ while $n \in \mathbb{N}$. The matrix representation of the reduced EP is tridiagonal in either parity basis and its determinant can be achieved by a three-term recursive equation similar to the one in Eq.~\eqref{eq_recu}. Once again a truncation of the basis is required in order to achieve a characteristic equation whose roots are the eigenenergies of the Rabi model. The errors then stems from the truncated basis and can be reduced by increasing the size of the basis. The possible excluded solutions can be searched for through examining $H_{R}|\Psi \rangle=(n-\Delta) |\Psi \rangle$.

Other spin-boson models can also be treated in the same way. Although the tridiagonality of the matrix representation facilitates the construction of the determinant, it is not essential in solving the problem. The difficulty arises in finding the roots of the characteristic equation where the graphical root-finding approach is adopted in Rabi-type models \cite{Xie}, similar to the plot in Fig.~\ref{GR1}.

\section{Many-spin examples}\label{sec_MSE}
In this section, the USUSY QM of single spins are exploited in some models with more than one spin. Examples include a model of spin-spin interactions, TC model and generalized Dicke models. Specially here we interpret the solutions of the TC model from a MPS point of view.

\subsection{A model of spin-spin interactions}\label{susec_2S}
\begin{figure}[tbh]
\begin{centering}
\begin{tikzpicture}
\begin{scope}[xshift=0.00\textwidth,yshift=0.23\textwidth]
\node[anchor=south west,inner sep=0](image) at (0,0){
\includegraphics[width=0.4\textwidth]{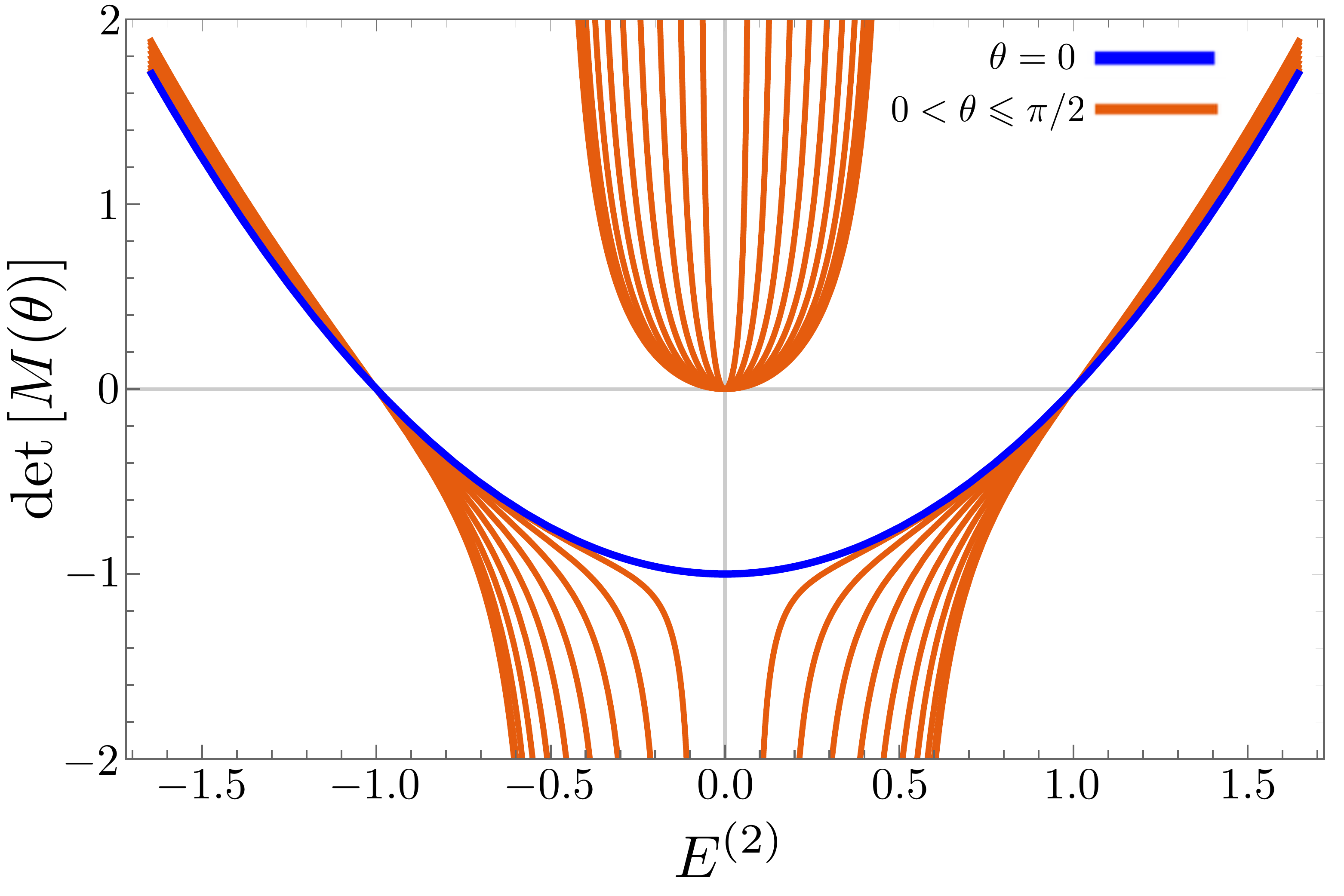}};
\end{scope}
\end{tikzpicture}
\par\end{centering}
\caption{(color online) Plots of $\det{[M(\theta)]}$ for different $\theta$. The characteristic equation of the original reduced Hamiltonian $\theta=0$, shown by the blue line, excludes two zero-energy solutions of $H^{(2)}$ while the characteristic equations corresponding to the transformed Hamiltonian with $\theta  \neq n \pi$ for $n \in \mathbb{Z}$ include all the solutions.}
\label{2S}
\end{figure}

We first solve the EP of two interacting spins
\begin{equation}
	H^{(2)}=\sigma_{+}^{(1)}\sigma_{-}^{(2)}+h.c., \quad  H^{(2)}|\Psi^{(2)} \rangle=E^{(2)} |\Psi^{(2)} \rangle 
\end{equation}
in order to exemplify how the maximal information can be extracted from the reduced EPs. Next, we explain how the achieved eigensolutions can be exploited to solve the reduced EPs of a three-spin system. The procedure can be iteratively carried on to include more spins one by one and construct larger systems. A remark on the notation is that the superscript of the Hamiltonians and corresponding eigenvalues/eigenstates denotes how many spins are included in them while the superscript of the spin operators labels the single spin. Representing $H^{(2)}$ in eigenspinors of $\sigma_{z}^{(2)}$
\begin{equation}
	H^{(2)}=
	\begin{pmatrix}
      0&  \sigma_{-}^{(1)} \\
      \sigma_{+}^{(1)}   &0
\end{pmatrix},
\end{equation}
the associated reduced Hamiltonians are
\begin{equation}
	h^{(1)}_{\pm}=\frac{1}{E^{(2)}} \sigma_{\mp}^{(1)}\sigma_{\pm}^{(1)}.
\end{equation}
The square of $H^{(2)}$ has a standard SUSY QM, however, we follow to analyze the associated USUSY QM. $H^{(2)}$ has four eigensolutions: three triplet states
\begin{eqnarray}
	&&|\Psi_{1}^{(2)} \rangle =\mid--\rangle, \,|\Psi_{2}^{(2)} \rangle =\mid++\rangle,\nonumber \\
	&&|\Psi_{3}^{(2)} \rangle = \frac{1}{\sqrt{2}} \left(\mid+-\rangle+\mid-+\rangle  \right) 
\end{eqnarray}
with energies $E^{(2)}_{1,2}=0$, $E^{(2)}_{3}=1$ and one singlet state
\begin{equation}
|\Psi_{4}^{(2)} \rangle =  \frac{1}{\sqrt{2}} \left(\mid+-\rangle-\mid-+\rangle  \right) 
\end{equation}
with energy $E^{(2)}_{4}=-1$, where the local basis is chosen as $\sigma_{z}^{(k)} |\pm \rangle=\pm |\pm \rangle$. Therefore, $h^{(1)}_{\pm}$ do not exist for states with a zero energy, and hence, two zero-energy states would be missing from the solutions of either reduced Hamiltonian. We note that zero-energy eigenstates are also disentangled states. If the two-spin Hamiltonian is rotated by
\begin{equation}
	U \equiv U(\theta)=e^{i \sigma^{(2)}_{y} \theta/2 }, \quad 0< \theta <\pi,
\end{equation}
we find
\begin{equation}
	\tilde{H}^{(2)}= U^{\dag} H^{(2)}U=
	\begin{pmatrix}
      -\tilde{H}^{(1)}&  \tilde{F}_{+} \\
      \tilde{F}_{-}   &\tilde{H}^{(1)}
\end{pmatrix},
\label{eq_TrH}
\end{equation}
where
\begin{eqnarray}
	\tilde{H}^{(1)}&=&\frac{1}{2} \sin{\theta}  \sigma_{x}^{(1)} , \\ \tilde{F}_{\pm}&=& \sigma_{\mp}^{(1)} \cos^{2}{(\frac{\theta}{2})}-\sigma_{\pm}^{(1)} \sin^{2}{(\frac{\theta}{2})}.
\end{eqnarray}
Expanding the reduced EP $(\tilde{h}^{(1)}_{+}-E^{(2)} )|\tilde{\psi}_{+} \rangle=0$ in the eigenbasis of $\tilde{H}^{(1)}$ and showing its matrix representation by $M(\theta)$, eigenenergies of $\tilde{h}^{(1)}_{+}$ are zeros of $\det{[M(\theta)]}$ as a function of $E^{(2)}$. The eigenenergies of the reduced Hamiltonian cover now the full eigenenergies of $H^{(2)}$ if $\theta $ is not an integer multiple of $\pi$, see Fig.~\ref{2S}. The drawback of the transformation is that the reduced Hamiltonians $\tilde{h}^{(1)}_{\pm}$ do not have the conserved quantity of the initial reduced Hamiltonians $[h^{(1)}_{\pm},\sigma_{z}^{(1)}]=0$. Clearly, the good quantum number of the full Hamiltonian is maintained after the transformation
\begin{equation}
	U^{\dag}[H^{(2)},\sigma_{z}^{(1)}+\sigma_{z}^{(2)}]U=[\tilde{H}^{(2)},\sigma_{z}^{(1)}+U^{\dag}\sigma_{z}^{(2)}U]=0.
\end{equation}

The same situation recurs if we extend the model to include one extra spin, for instance,
\begin{equation}
	H^{(3)}=H^{(2)}+ \left(\tau \sigma_{+}^{(1)}\sigma_{-}^{(3)}+\sigma_{+}^{(2)}\sigma_{-}^{(3)}+h.c. \right),
\end{equation}
where $\tau $ is chosen to be a non-zero real number for the sake of simplicity. $H^{(3)}$ has the following form in the eigenspinors of $\sigma_{z}^{(3)}$
\begin{equation}
	H^{(3)}
	=
	\begin{pmatrix}
      H^{(2)}&   F(\tau) \\
      F^{\dag}(\tau)   &H^{(2)}
\end{pmatrix},\quad F(\tau)=\tau \sigma_{-}^{(1)}+ \sigma_{-}^{(2)}.
\end{equation}
$H^{(3)}$ has two disentangled eigenstates $\mid\pm\pm\pm\rangle$ as far as $|\tau|\neq1$. $h_{+}^{(2)}$ does not exist for $\mid---\rangle$ and the solution of this state would be missing from the solutions of $h_{+}^{(2)}$. Other eigenenergies of $H^{(3)}$ are captured by the reduced Hamiltonian. $h_{+}^{(2)}$ commutes with the $z$-component of total angular momentum and the eigenstates can be labeled by its eigenvalues
\begin{equation}
\sigma^{T}_{z}=\sigma_{z}^{(1)}+\sigma_{z}^{(2)}, \quad \sigma^{T}_{z} |\psi^{(2)}_{+} \rangle = m_{z}|\psi^{(2)}_{+} \rangle .
\end{equation}
The characteristic equation for the eigenstates of $h_{+}^{(2)}$ with $ m_{z}=-2$ is
\begin{equation}
	\frac{(1+\tau)^{2}}{E^{(3)}-1}+\frac{(1-\tau)^{2}}{E^{(3)}+1}-E^{(3)}=0.
	\label{eq_CEm2}
\end{equation}
which gives three eigenenergies
\begin{eqnarray}
	&&E_{1}^{(3)}=-\tau, \quad E_{2}^{(3)}=\frac{1}{2} \left(\tau-\sqrt{8+\tau^{2}} \right), \nonumber \\ &&E_{3}^{(3)}=\frac{1}{2} \left(\tau+\sqrt{8+\tau^{2}} \right). 
	\label{eq_m2sols}
\end{eqnarray}
There exists one state with zero-energy $E_{4}^{(2)}=0$ which has $ m_{z}=2$. The characteristic equation for the eigenstates of $h_{+}^{(2)}$ with $ m_{z}=0$ needs to be constructed by taking the determinant
\begin{equation}
	\begin{vmatrix}
 1-E+\frac{(1+\tau)^{2}}{2E}&\frac{1-\tau^{2}}{2E} \\ 
 \frac{1-\tau^{2}}{2E}& -1-E+\frac{(1-\tau)^{2}}{2E}
\end{vmatrix}=0,
\end{equation}
which renders the same set of solutions as those in Eq.~\eqref{eq_m2sols}. At $\tau=1$ for instance, eigenenergies of $H^{(3)}$ become $-1,0,2$, which are four-, two-, two-fold degenerate, respectively. Only one of the solutions with energy $-1$ is missing from the equation in Eq.~\eqref{eq_CEm2} which reduces the number of solutions of $h_{+}^{(2)}$ to six. In order to find out the excluded state, we consider
\begin{equation}
	F(\tau) |\Psi_{4}^{(2)} \rangle =\frac{\tau-1}{\sqrt{2}} |\Psi_{1}^{(2)} \rangle, 
\end{equation}
which vanishes if $\tau =1$ and allows to construct an eigenstate of $H^{(3)}$ as $ | - \rangle \otimes |\Psi_{4}^{(2)} \rangle $ in which the subsystem of spin $1,2$ is disentangled from the third spin. Also we note that the other eigenstates of $H^{(3)}$ with energy $-1$ do not fulfil $\langle \Psi_{4}^{(2)}|\psi_{-}^{(2)} \rangle \neq 0$ and so their solutions are not excluded. The degeneracy between $H^{(3)}$ and $H^{(2)}$ can be lifted by a $SU(2)$ transformation on the third spin. Subsequently, the eigensolutions of the associated reduced Hamiltonian cover all the solutions of $H^{(3)}$. In brief, the full Hamiltonian needs to be transformed in order to find a representation in which the components of the eigenstates are intertwined from one way e.g. 
\begin{equation}
|\tilde{\psi}_{-}^{(2)} \rangle = \frac{1}{E-\tilde{H}^{(2)}}\tilde{F}^{\dag}(\tau) |\tilde{\psi}_{+}^{(2)} \rangle,
\end{equation}
see the discussion at the end of Sec.~\ref{subsec_half}.

The procedure can be iterated in order to include more spins and achieve a targeted system. The iteration cannot run forever as at some stage the dimension of the Hilbert space exceeds the available resources required to solve the model. At this point, truncating the basis is required which allows to achieve a low-energy description of the model. The decimation procedure is elaborated in methods like the numerical renormalization group (NRG) \cite{Willson} or the density matrix renormalization group (DMRG) \cite{White,MPS01}. The common ground between the formalism presented here and NRG and DMRG is the essential role of the subsystems in constructing the full-system solution. DMRG minimizes the loss of information of an approximated solution by finding a truncated basis which maximizes the entanglement entropy \cite{Eisert,MPS01} of the wavefunction. A similar variational approach can be combined with the scheme of USUSY QM to overcome the limitations of the ED versus the growth of the Hilbert space with number of spins.
\subsection{Tavis-Cummings model}\label{susec_TC}
TC model \cite{TC} describes interactions of a single boson mode and $N$ spin-$\frac{1}{2}$. The TC Hamiltonian $H_{\scriptscriptstyle  TC}^{ \scriptscriptstyle (N)}$ is achieved by replacing $H_{  0}=a^{\dag} a$ and $ F^{\scriptscriptstyle (k)}_{+}=\alpha a$ in Eq.~\eqref{eq_HN01} for $k=1,\cdots, N$ although the following treatment works as well if $ F^{\scriptscriptstyle (k)}_{+}=\alpha^{\scriptscriptstyle (k)} a$ with a site-dependent $\alpha^{\scriptscriptstyle (k)}$. All the subsystems of the TC model need to be solved in order to construct the inner products in Eq.~\eqref{eq_MPC02} and subsequently a MPS representation similar to the one in Eq.~\eqref{eq_NSS}. However, the symmetry of the TC model
\begin{equation}
\left[H^{\scriptscriptstyle(N)}_{\scriptscriptstyle TC},a^{\dag}a+\frac{1}{2} \sum_{k=1}^{N}\sigma_{z}^{\scriptscriptstyle (k)}\right]=0,
\end{equation}
allows to construct a MPS ansatz $|\Psi^{\scriptscriptstyle {  (N)}}_{  n,\eta} \rangle= \mathcal{M}^{  \scriptscriptstyle(N)}_{  n} |\varsigma^{}_{  n,\eta} \rangle$ with 
\begin{equation}
	\mathcal{M}^{\scriptscriptstyle  (N)}_{  n}
	=
	\prod_{  \ell=0}^{  N-1} I_{  \ell} \otimes 
	\begin{pmatrix}
   a I_{  N-\ell-1}   &  0  \\
     0 &   I_{  N-\ell-1}
     \end{pmatrix}
	\,I_{  N}|n \rangle, 
	\label{eq_TCMPS01}
\end{equation}
which imitates the entanglement pattern of the state in Eq.~\eqref{eq_NSS} and specifically Eq.~\eqref{eq_NSS02}. $I_{\ell}$ is $2^{\ell} \times 2^{\ell}$ identity matrix, $|\varsigma^{}_{  n,\eta} \rangle$ is a spinor with $2^{ \scriptscriptstyle N}$ unknown parameters as its components, $n$ stands for the boson occupation number while $\eta \in \{1,\cdots, 2^{N} \}$ is a quantum number associated with spin part. Because $\mathcal{M}^{  \scriptscriptstyle(N)^{\dag}}_{n^{\prime}} H_{ \scriptscriptstyle TC}^{ \scriptscriptstyle (N)} \mathcal{M}^{ \scriptscriptstyle (N)}_{n}=0$ for $n\neq n^{\prime}$, the boson DOF is fixed by the ansatz and the problem is reduced to determining $|\varsigma_{  n,\eta} \rangle$.

The eigensolutions can be grouped into two classes regarding $n \geqslant N$ or $n < N$. In the first class, the associated eigenenergies are roots of the determinant of $\mathcal{M}^{\scriptscriptstyle{  (N)}^{\dag}}_{n}(H_{  TC}^{\scriptscriptstyle  (N)}-E) \mathcal{M}^{\scriptscriptstyle  (N)}_{n}$ and $2^{N}$ parameters need to be calculated. In the second class, the states have less parameters to be determined. For instance, no parameter needs to be found for the disentangled state with $n=0$ and with the eigenenergy $-N \Delta$, where the eigenstate is denoted by $|\Psi^{\scriptscriptstyle {  (N)}}_{  0} \rangle$. The number of unknown coefficients are $N+1 $ and $N(N+1)/2+1$ for $n=1$ and $n=2$, respectively. Generally, the number of parameters is determined by
\begin{equation}
	P(N,n)=\sum_{j=0}^{n} \begin{pmatrix}
      N   \\
        j
\end{pmatrix},
\label{eq_dis}
\end{equation}
see App.~\ref{app_TCMPS} for a demonstration. It can be readily shown that there is an upper bound to the number of parameters $P(N,n) \leqslant (1+N)^{n}$ for arbitrary $n$. Hence, if $N \gg n$, then $P(N,n) \sim N^{n}$ which shows the number of parameters increases polynomially with $N$.

If the spin and the fields are not correlated strongly $\alpha < \Delta<1$, the ground state is in the second class: assume $|\alpha|$ is increased adiabatically from zero for $\Delta>0$, then the ground state is shifted from $|\Psi^{\scriptscriptstyle {  (N)}}_{  0} \rangle$ to one of the states in $|\Psi^{ \scriptscriptstyle{  (N)}}_{  1,\eta} \rangle$ and then to one of the states in $|\Psi^{ \scriptscriptstyle{  (N)}}_{  2,\eta} \rangle$ and so on. 

\subsection{Generalized Dicke model}\label{susec_gD}

\begin{figure}[tb]
\begin{centering}
\begin{tikzpicture}
\begin{scope}[xshift=0.00\textwidth,yshift=0.23\textwidth]
\node[anchor=south west,inner sep=0](image) at (0,0){
\includegraphics[width=0.48\textwidth]{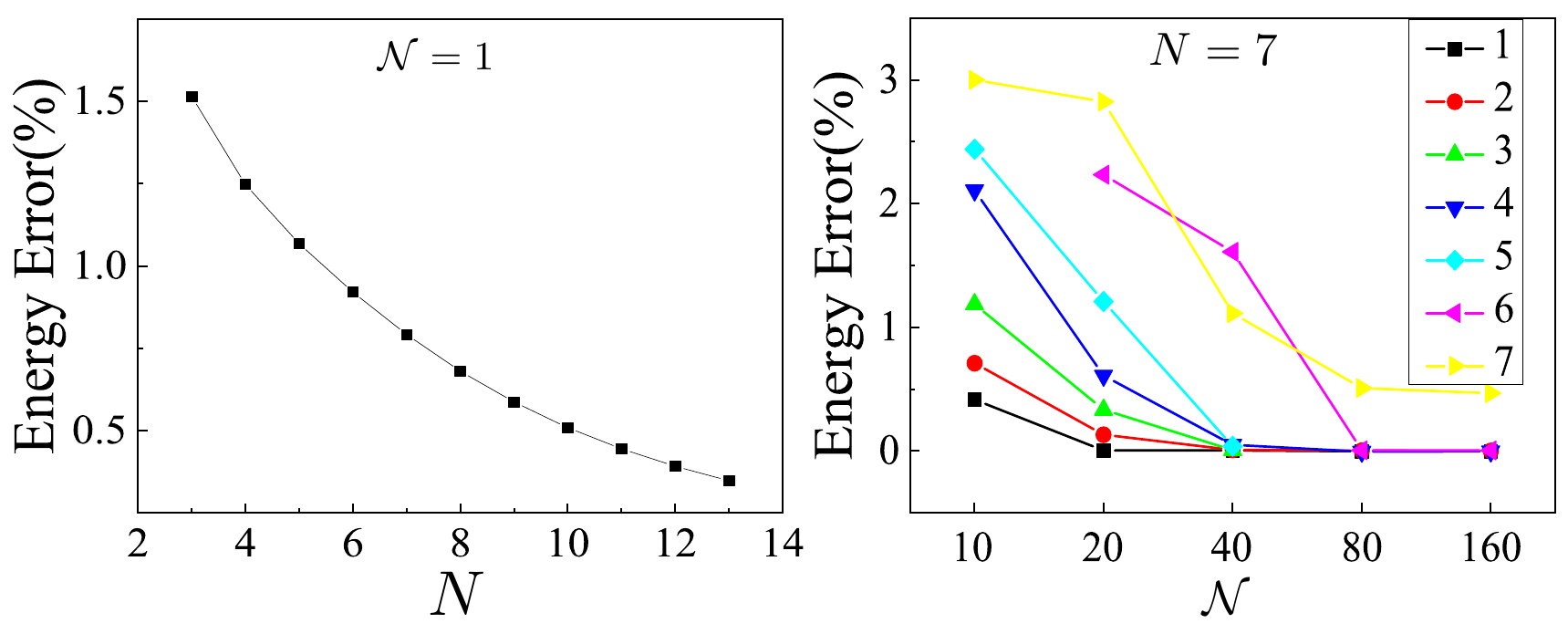}};
\end{scope}
\end{tikzpicture}
\par\end{centering}
\caption{(color online) Results of genralized Dicke models with $\alpha=0.15$, $\beta=0.1$, $\gamma=0.1$ and $\Delta=0.2$. Models are solved via diagonalizing the reduced Hamiltonian in a $\mathcal{N}$-dimensional basis and then outcomes are compared with results of a direct ED engaging a $2^{N} \times M$ dimensional basis, where the number of lowest boson states is fixed by $M=20$. Left: deviations in the ground-state energy versus number of spins found by $\mathcal{N}=1$. Right: deviations in seven distinct low-lying eigenenergies versus $\mathcal{N}$ for a model with $N=7$. The results of each eigenstate are connected by a line. The numbers in the legend from $1$ to $7$ stand for low to high energies, respectively.}
\label{GR2}
\end{figure}
The last example is the generalized Dicke model \cite{Dicke,Braak_2013,He_2015,Kirton} characterized by replacing $H_{  0}=a^{\dag} a$ and $F^{(k)}_{+}=\alpha a+\beta a^{\dag}+\gamma$ in Eq.~\eqref{eq_HN01} for $k=1,\cdots, N$. In Fig.~\ref{GR2}, we show the results of the EP of $N$-spin model through diagonalizing the associated reduced Hamiltonian $h_{ -}^{\scriptscriptstyle{  (N-1)}}$. The basis for representing $h_{  -}^{\scriptscriptstyle{  (N-1)}}$ is constructed out of $\mathcal{N}$ lowest eigenstates of $H^{  \scriptscriptstyle(N-1)}_{  +}$. The results are compared with outcomes of a direct ED on $H^{  \scriptscriptstyle(N)}$ by means of a basis with a dimension of $2^{N} \times M$. It is the merits of the intrinsic correlated basis that even by shrinking the size of the truncated basis down to $\mathcal{N}=1$, the ground-state energy can be achieved quite accurately in comparison with the direct ED, see also App.~\ref{app_EDD}. In this case, the equation
\begin{equation}
	\langle \Psi^{ \scriptscriptstyle(N-1)}_{0} |h_{-}| \Psi^{ \scriptscriptstyle(N-1)}_{0} \rangle \approx E,
	\label{eq_expv}
\end{equation}
where the expectation value is taken by the ground state of $H_{+}^{\scriptscriptstyle(N-1)}$, needs to be solved for $E$ and the smallest solution is a candidate for the ground state of $H^{\scriptscriptstyle(N)}$. We note that no optimization in Eq.~\eqref{eq_expv} needs to be performed and the source of errors is the truncated basis. As it can be seen in Fig.~\ref{GR2}, the results become more accurate by increasing the number of spins $N$. The latter can be attributed to the pattern of correlations in Eq.~\eqref{eq_NSS02}: the higher number of spins induces more correlations into characteristic equations.

\section{Discussion}
The EP of a Hamiltonian describing interactions between a single spin-$\frac{p}{2}$ and external fields provides intertwining relationships between the components of the eigenstate. The existence of these correlations implies there exist degrees of redundancy in fixing the quantum state of the system. In this work, we exploited these relations to reduce DOF required to solve the EP and also to derive USUSY QM. The formalism reveals that a subsystem of the system can carry all information encoded in the full eigenstate. The scheme is non-trivial if the Hamiltonian does not bear a good spin quantum number. In such cases the eigenstates are expected to be entangled states. It can be concluded that the entanglement provides correlations between constituents of the system and can be exercised to reduce DOF in spin models. It is curious to find out whether the same conclusion can be made in a wider class of interacting problems or not which we leave it for future studies. 

The importance of the entanglement in exploring relevant sectors of Hilbert spaces is elevated in tensor network (TN) methods \cite{Verstraete,ORUS,Vidal02} developed under the influence of the quantum information science. A TN state is a trial wavefunction which allows to control the amount of the entanglement in the state. 
Our study can also be of relevance to this realm since it reveals that a system of interacting spins offers its own intrinsic basis with a tractable pattern of entanglement that can be deduced directly from the Hamiltonian. The connection becomes tangible by noticing that the generic representation of a many-spin eigenstate in Eq.~\eqref{eq_NSS02} is a MPS \cite{MPS01}, which is the simplest TN state. Moreover, the fractal structure of the states in Eq.~\eqref{eq_NSS02} mimics the way by which entanglement is implemented in some TN ansatz \cite{Vidal01,Vidal03}. It can be subjects of further researches to connect the study presented here and TN methods in order to construct new families of ansatz in the study of spin systems.

\section*{Acknowledgment}
We thank Reinhold Egger for useful discussions. This work is supported by NSF-China under Grant No. 11804396, and is carried out in part using computing resources at High Performance Computing Centre of Central South University.
\appendix
\renewcommand\thefigure{A\arabic{figure}} \setcounter{figure}{0}

\section{Unitary Transformations of Generalized Rabi Models}\label{app_UT}
We consider intially a $2\times 2$ matrix Hamiltonian and its transformation under a rotation in $SU(2)$
\begin{equation}
	H
	=
	\begin{pmatrix}
     H_{+} &F_{+}    \\
    F_{-} &H_{-}  
\end{pmatrix},
\quad
	\tilde{H}
	=
	\begin{pmatrix}
     \tilde{H}_{+} &\tilde{F}_{+}    \\
    \tilde{F}_{-} &\tilde{H}_{-}  
\end{pmatrix}
	= U^{\dag} H U,
\end{equation}
where
\begin{equation}
	U^{}
	\equiv
	U^{}(\phi,\theta)
	=
	\begin{pmatrix}
     e^{i \phi} \cos{\frac{\theta}{2}} &e^{i \phi} \sin{\frac{\theta}{2}}    \\
     \\
    -e^{-i \phi} \sin{\frac{\theta}{2}}&e^{-i \phi} \cos{\frac{\theta}{2}}  
\end{pmatrix}.
\end{equation}
Given $H_{\pm}=H_{0}\pm \Delta$, the explicit form of the transformed Hamiltonian is written in terms of the following operators
\begin{eqnarray}
	  \tilde{H}_{\pm} &=&
	  H_{0}\pm \Delta \cos{\theta} \mp 
	  \frac{1}{2} 
	   \left(  e^{i \phi} F_{-}+e^{-i \phi} F_{+}  \right)
	  \sin{\theta},\\
	  \tilde{F}_{+}
	  &=&
	  \Delta \sin{\theta}+
	  F_{+} e^{-i \phi}\cos^{2}{\frac{\theta}{2}} 
	  -F_{-} e^{i \phi}\sin^{2}{\frac{\theta}{2}} .
\end{eqnarray}
We proceed with the generalized Rabi model specified by $H_{0}=a^{\dag}a$ and $F_{+}=\alpha a+\beta a^{\dag} +\gamma$ for $\alpha,\beta \in \mathbb{R}^{+}$ and $\gamma \in \mathbb{C}$. The entries of the transformed Hamiltonian are
\begin{eqnarray}
\tilde{H}_{\pm}& =&a^{\dag}a \pm \left( a \lambda^{\ast}+a^{\dag} \lambda +\tilde{\Delta}\right),\\ \tilde{F}_{+}&=&\tilde{\alpha} a+\tilde{\beta} a^{\dag}+\tilde{\gamma}=\tilde{F}_{-}^{\dag}
\end{eqnarray}
where
\begin{eqnarray}
	\lambda&=&-\frac{1}{2}\sin{\theta} \left(\alpha e^{i \phi}+\beta e^{-i\phi} \right),
	\\
	\tilde{\Delta}&=&\Delta \cos{\theta}-\frac{1}{2} \sin{\theta} \left(\gamma e^{-i \phi}+\gamma^{\ast} e^{i\phi} \right),
	\\
	 \tilde{\alpha}&=&\alpha e^{-i \phi} \cos^{2}{\frac{\theta}{2}}  -\beta e^{i \phi} \sin^{2}{\frac{\theta}{2}} , 
	 \\
	 \tilde{\beta}&=&\beta e^{-i \phi} \cos^{2}{\frac{\theta}{2}}  -\alpha e^{i \phi} \sin^{2}{\frac{\theta}{2}} ,
	 \\
	  \tilde{\gamma}&=&\Delta \sin{\theta}+\gamma e^{-i \phi} \cos^{2}{\frac{\theta}{2}}  -\gamma^{\ast} e^{i \phi} \sin^{2}{\frac{\theta}{2}} .
\end{eqnarray}
$\tilde{H}_{\pm}$ satisfy eigenvalue problems $\tilde{H}_{\pm} |\tilde{n};\pm \rangle = \tilde{E}^{(n)}_{\pm}|\tilde{n};\pm \rangle$, where $|\tilde{n};\pm \rangle=D^{\dag}(\pm \lambda) |n \rangle$ is defined by action of the displacement operator $D^{}( \lambda)=e^{\lambda a^{\dag}-\lambda^{\ast} a}$ on the eigenket of the number operator $a^{\dag} a |n \rangle=n |n \rangle$, and the eigenenergies are
\begin{equation}
\tilde{E}^{(n)}_{\pm}=n-|\lambda|^{2}\pm \tilde{\Delta}.
\end{equation}
The commutation relations between the fields in the Hamiltonian change as
\begin{eqnarray}
	\left[\tilde{H}_{\pm}, \tilde{F}_{+} \right]&=&\tilde{\beta}( a^{\dag}\pm\lambda^{\ast}  )-\tilde{\alpha} (a\pm \lambda)
	\label{eq_commu01}
	,
	\\
	\left[\tilde{F}_{+}, \tilde{F}_{-} \right]&=&|\tilde{\alpha}|^{2}-|\tilde{\beta}|^{2}=\left( {\alpha}^{2}-{\beta}^{2}\right) \cos{\theta},
	\\
	\left[\tilde{H}_{+}, \tilde{H}_{-} \right]&=&2 \left( \lambda^{\ast}a-\lambda a^{\dag} \right).
\end{eqnarray}

\section{Tridiagonalizing Matrix Representations of Generalized Rabi Models}\label{app_TMRGR}
Assume neither of $\alpha, \beta$ and $\gamma$ is zero. Based on the commutation relation in Eq.~\eqref{eq_commu01}, setting either $\tilde{\alpha}=0$ or $\tilde{\beta}=0$ turns the representation of reduced Hamiltonians 
\begin{equation}
	\tilde{h}_{\pm}=\tilde{H}_{\pm}+\tilde{F}_{\pm} \tilde{G}_{\mp} \tilde{F}_{\mp}, \quad \tilde{G}_{\mp}=\frac{1}{E-\tilde{H}_{\mp}} ,
\end{equation}
into a tridiagonal form when represented in the basis of $\tilde{H}_{\mp}$. To illustrate it, we choose $\tilde{\beta}=0$ which is satisfied by $\tan{\frac{\theta}{2}}=\sqrt{\beta/\alpha}$ and $\phi=0$. Concerning the eigenvalue problem $(\tilde{h}_{+} -E)|\tilde{\psi}_{+}\rangle=0$, the matrix elements in the base ket of $|\tilde{n};- \rangle$ become
\begin{eqnarray}
	\tilde{f}^{(m,n)}&=&\langle \tilde{m};-|(\tilde{h}_{+} -E)|\tilde{n};- \rangle
	\nonumber
	\\ 
	&=&
	\left[m+3|\lambda|^{2}+\tilde{\Delta}+(m+1)|\tilde{\alpha}|^{2}\tilde{\epsilon}_{-}^{(m+1)} \right.
	\nonumber
	\\ 
	&&
	\left.+|\tilde{\gamma}+\tilde{\alpha} \lambda|^{2}\tilde{\epsilon}_{-}^{(m)}-E
	\right]\delta_{m,n}
	\nonumber
	\\
	&&+
	\left[
	2 \lambda^{\ast}+\tilde{\alpha} (\tilde{\gamma}^{\ast}+\tilde{\alpha}^{\ast} \lambda^{\ast}) \tilde{\epsilon}_{-}^{(n)}
	\right] \sqrt{n} \delta_{m+1,n}
	\nonumber
	\\ 
	&&+
	\left[
	2 \lambda+\tilde{\alpha}^{\ast} (\tilde{\gamma}+\tilde{\alpha} \lambda) \tilde{\epsilon}_{-}^{(m)}
	\right] \sqrt{m} \delta_{m,n+1},
\end{eqnarray}
where $\tilde{\epsilon}_{-}^{(m)}=( E-\tilde{E}^{(m)}_{-})^{-1}$. The determinant of the tridiagonal matrix can be worked out by means of the three-term recurrence relation in Eq.~\eqref{eq_recu}.

\section{Exact Isolated Solutions of Genralized Rabi Models}\label{app_IES}
We consider three classes of exact isolated solutions according to the commutation relations between $\tilde{H}_{\pm}$, $\tilde{F}_{+}$ and $\tilde{F}_{-}$ for non-zero $\alpha$ and $\beta$. We set $\phi=0$ and treat $\theta$ as a free parameter to be fixed restricted to $0< \theta <\pi $. The USUSY QM does not exist for the states with the exact solution. Two first classes are about genralized Rabi models $\alpha \neq \beta$ and the third class is concerned with the Rabi model $\alpha =\beta$.

\subsection{First Class}
If $\tilde{\beta} =0$ and $\lambda \tilde{\alpha}=-\tilde{\gamma}$, then $[\tilde{H}_{-}, \tilde{F}_{+}]=-\tilde{F}_{+}$. That is $\tilde{F}_{+}$ is the annihilation operator in an algebra $[\tilde{F}_{+}, \tilde{F}_{-} ]=|\tilde{\alpha}|^{2}>0$ diagonalizing the Hamiltonian $\tilde{H}_{-}=\tilde{F}_{-}\tilde{F}_{+}+const.$. The constraints read explicitly as
\begin{eqnarray}
\begin{cases}
      \tan{\frac{\theta}{2}}=\eta \sqrt{\frac{\beta}{\alpha}} &
       \text{ for } \eta=\pm   ,
      \\
     \frac{ \alpha^{2}-\beta^{2}}{2}= \Delta + \gamma   \cot{\theta} & \text{ for } \gamma \in\mathbb{R}.
\end{cases}
	\label{eq_ISO01}
\end{eqnarray} 
Two solutions of $\theta$ in the limited range are labeled by $\eta=\pm$. For either solution, if $|\tilde{0} \rangle$ is a vacuum of $\tilde{F}_{-}  $, then $\langle \tilde{\Psi}_{} |=(0\,\,\,\langle \tilde{0}|)$ is an eigenstate of the transformed Hamiltonian $\tilde{H}$. The state $|\tilde{\Psi}  \rangle$ is disentangled and does not bear the USUSY QM since one of the supercharges does not exist. We note that if $\gamma =0$, two disentangled eigenstates are degenerate and therefore their superposition can entangle the boson state and the spin state. 

If $\tilde{\alpha}=0$ and $\lambda \tilde{\beta}=\tilde{\gamma}$, then $[\tilde{H}_{+}, \tilde{F}_{-}]=-\tilde{F}_{-}$. That is $\tilde{F}_{-}$ is the annihilation operator in an algebra $[\tilde{F}_{+}, \tilde{F}_{-} ]=-|\tilde{\beta}|^{2}<0$ diagonalizing the Hamiltonian $\tilde{H}_{+}=\tilde{F}_{+}\tilde{F}_{-}+const.$. The solutions in this case can be achieved by a unitary transformation from the proceeding case. $4 \times 4$ supersymmetric Hamiltonians are constructed along the baseline of parameters in Eq.~\eqref{eq_ISO01} for JC model $\beta=\gamma=0$ in \cite{ANDREEV} and for generalized Rabi models with a parity symmetry $\gamma=0$ in \cite{Tomka:2015aa}.

\subsection{Second Class}
If $\tilde{\beta} =0$ and $\lambda \tilde{\alpha}=\tilde{\gamma}$, then $[\tilde{H}_{+}, \tilde{F}_{+}]=-\tilde{F}_{+}$. That is $\tilde{F}_{+}$ is the annihilation operator in an algebra diagonalizing the Hamiltonian $\tilde{H}_{+}=\tilde{F}_{-}\tilde{F}_{+}+const.$. The constraints on the parameters are
\begin{eqnarray}
\begin{cases}
      \tan{\frac{\theta}{2}}=\eta \sqrt{\frac{\beta}{\alpha}}
      &
       \text{ for } \eta=\pm   ,
      \\
     \frac{ \beta^{2}-\alpha^{2}}{2}= \Delta + \gamma   \cot{\theta} & \text{ for } \gamma \in\mathbb{R}.
\end{cases}
	\label{eq_ISO02}
\end{eqnarray} 
The anzatz
\begin{equation}
	|\tilde{\Psi}_{M} \rangle=
	\sum_{m=0}^{M}
	\begin{pmatrix}
     c^{(m)}_{+} | \tilde{m};+ \rangle    \\
     c^{(m)}_{-} | \tilde{m};+ \rangle  
\end{pmatrix}
\end{equation}
with $M\in \big\{0,1,2,\cdots \big\}$, where $ | \tilde{m};+ \rangle$ is the eigenstate of $\tilde{H}_{+}$ with the eigenenergy $\tilde{E}^{(m)}_{+}=m-|\lambda|^{2}+ \tilde{\Delta}$, leads to two coupled recurrence equations 
\begin{eqnarray}
   &&   c_{+}^{(p)} ( p+\tilde{\Delta}-\lambda^{2}-E )+c_{-}^{(p+1)} \tilde{\alpha} \sqrt{p+1}=0, \\
     && c_{-}^{(p)} ( p-\tilde{\Delta}+3 \lambda^{2}-E )-c_{-}^{(p+1)}2 \lambda  \sqrt{p+1}
      \nonumber \\
   &&   -c_{-}^{(p-1)}2 \lambda  \sqrt{p}+c_{+}^{(p-1)} \tilde{\alpha} \sqrt{p}=0,
\end{eqnarray}
by implementing it in $\tilde{H}|\tilde{\Psi}_{M} \rangle=E|\tilde{\Psi}_{M} \rangle$. The ansatz has $2(M+1)$ unknown coefficients $c_{\pm}^{(p)}$ for $0 \leqslant p \leqslant M$. Since the coefficients of the largest quantum number are linearly related $2 \lambda c_{-}^{(M)}=\tilde{\alpha} c_{+}^{(M)}$, the set of equations can be represented as a matrix equation of rank $2M+1$. The ansatz is an eigenstate of $\tilde{H}$ with $E=\tilde{E}^{(M)}_{+}$ if the determinant of the matrix is zero which introduces an extra constraint on the parameters of the model. For instance, one finds $2 \lambda^{2} =\tilde{\Delta}$ for $M=0$; and in the case of $M=1$, the condition is set by a determinant
\begin{eqnarray}
&&	\begin{pmatrix}
1 &0  & -\tilde{\alpha} \\ 
 0& 1+2 \tilde{\Delta}-4 \lambda^{2} &2 \lambda \\ 
 -\tilde{\alpha}&  2 \lambda & 2 \tilde{\Delta}-4 \lambda^{2}
\end{pmatrix}
\begin{pmatrix}
      c_{+}^{(0)}    \\
      c_{-}^{(0)}     \\
      c_{-}^{(1)}
\end{pmatrix}
=0
\nonumber
\\
&&\Rightarrow
	\begin{vmatrix}
1 &0  & -\tilde{\alpha} \\ 
 0& 1+2 \tilde{\Delta}-4 \lambda^{2} &2 \lambda \\ 
 -\tilde{\alpha}&  2 \lambda & 2 \tilde{\Delta}-4 \lambda^{2}
\end{vmatrix}=0.
\end{eqnarray}
The eigenstates are degenerate if $\gamma=0$ once again. Within an equivalent path, one can follow to find the solutions through setting $\tilde{\alpha}=0$ and $\lambda \tilde{\beta}=-\tilde{\gamma}$ which provide $[\tilde{H}_{-}, \tilde{F}_{-}]=-\tilde{F}_{-}$.

\subsection{Third Class}
In this class we deal with Rabi models $\alpha=\beta$ in the presence of an in-plane magnetic field. An ansatz 
\begin{equation}
	|\tilde{\Psi}_{M} \rangle=
	\begin{pmatrix}
     \sum_{m=0}^{M}c^{(m)}_{+} | \tilde{m};+ \rangle   \\
     \sum_{m=0}^{M-1} c^{(m)}_{-} | \tilde{m};+ \rangle  
\end{pmatrix}
\end{equation}
with $M\in \big\{1,2,3,\cdots \big\}$. Inserting $|\tilde{\Psi}_{M} \rangle$ in the EP, a system of equations can be derived. $|\tilde{\Psi}_{M} \rangle$ can be an eigenstate of $\tilde{H}$ with $E=\tilde{E}^{(M)}_{+}$ if the system of equations can be solved self-consistently. It can come true by setting $\tilde{\alpha}=\tilde{\beta}=0$ which fixes the angle of rotation as $\tan^{2}{\frac{\theta}{2}}=1$. To illustrate it, first, we consider $M=1$ by which the set of equations can have non-trivial solutions if a condition on the parameters can be met
\begin{eqnarray}
&&	\begin{pmatrix}
1   & -\tilde{\gamma} \\ 
 -\tilde{\gamma}^{\ast}& 1+2 \tilde{\Delta}-4 \lambda^{2} 
 \end{pmatrix}
\begin{pmatrix}
      c_{+}^{(0)}    \\
      c_{-}^{(0)}     
\end{pmatrix}
=0
\nonumber
\\
\Rightarrow
&&	\begin{vmatrix}
1   & -\tilde{\gamma} \\ 
 -\tilde{\gamma}^{\ast}& 1+2 \tilde{\Delta}-4 \lambda^{2} 
\end{vmatrix}=0,
\end{eqnarray}
and the coefficient with the largest quantum number is fixed by $\tilde{\gamma}^{\ast}c_{+}^{(1)} =2 \lambda c_{-}^{(0)}$. For $M=2$, one finds
\begin{eqnarray}
&&	\begin{pmatrix}
2 & -\tilde{\gamma}&0&0 \\ 
 -\tilde{\gamma}^{\ast}& 2(1+ \tilde{\Delta}-2 \lambda^{2})&0 &2 \lambda \\ 
0&0&1& -\tilde{\gamma}\\
0&2 \lambda&-\tilde{\gamma}^{\ast}& 1+ 2\tilde{\Delta}-4 \lambda^{2} 
\end{pmatrix}
\begin{pmatrix}
      c_{+}^{(0)}    \\
      c_{-}^{(0)}     \\
      c_{+}^{(1)}     \\
      c_{-}^{(1)}
\end{pmatrix}
=0
\nonumber
\\
&& \Rightarrow
	\begin{vmatrix}
2 & -\tilde{\gamma}&0&0 \\ 
 -\tilde{\gamma}^{\ast}& 2(1+ \tilde{\Delta}-2 \lambda^{2})&0 &2 \lambda \\ 
0&0&1& -\tilde{\gamma}\\
0&2 \lambda&-\tilde{\gamma}^{\ast}& 1+ 2\tilde{\Delta}-4 \lambda^{2} 
\end{vmatrix}=0,
\end{eqnarray}
accompanied with $\tilde{\gamma}^{\ast}c_{+}^{(2)} =2 \sqrt{2}\lambda c_{-}^{(1)}$. Here also the states are degenerate if $\gamma=0$ as two solutions of $\theta$ do not generate distinct models.

\section{Diagonalizing a spin-$1$ model in the spin subspace}\label{app_S1}
Hamiltonian of a single spin-$1$ coupled with an external field in the associated eigenvalue problem reads
\begin{equation}
	\begin{pmatrix}
      H_{0}+ \Delta &F_{+}&0    \\
     F_{-} & H_{0}  &  F_{+} \\
     0&F_{-}&H_{0}- \Delta  
\end{pmatrix}
\begin{pmatrix}
     |\psi_{1} \rangle\\
     |\psi_{2} \rangle\\
    |\psi_{3} \rangle 
\end{pmatrix}
	=
	E
	\begin{pmatrix}
     |\psi_{1} \rangle\\
     |\psi_{2} \rangle\\
    |\psi_{3} \rangle 
\end{pmatrix}.
\end{equation}
Making the Hamiltonian traceless, we find
\begin{equation}
	\begin{pmatrix}
      0 &G_{1} F_{+}&0    \\
      G_{2} F_{-} & 0  &   G_{2} F_{+} \\
     0& G_{3} F_{-}&0  
\end{pmatrix}
\begin{pmatrix}
     |\psi_{1} \rangle\\
     |\psi_{2} \rangle\\
    |\psi_{3} \rangle 
\end{pmatrix}
	=
	\begin{pmatrix}
     |\psi_{1} \rangle\\
     |\psi_{2} \rangle\\
    |\psi_{3} \rangle 
\end{pmatrix},
\end{equation}
where
\begin{equation}
\begin{cases}
     G_{1}=\left( E-H_{0}-\Delta \right)^{-1}, \\
     G_{2}=\left( E-H_{0} \right)^{-1},\\
     G_{3}=\left( E-H_{0}+\Delta \right)^{-1}.
\end{cases}
\end{equation}
Taking the square of the traceless matrix decouples one of the components from two others
\begin{equation}
	\left( H_{0}+F_{-}G_{1}F_{+}+F_{+}G_{3}F_{-} \right) |\psi_{2} \rangle =E |\psi_{2} \rangle.
	\label{eq_SP_2}
\end{equation} 
Two other components are coupled by a $2 \times 2$ matrix
\begin{equation}
\begin{pmatrix}
      G_{1} F_{+} G_{2} F_{-}& G_{1} F_{+}  G_{2} F_{+}     \\
      G_{3} F_{-} G_{2} F_{-} &G_{3} F_{-}  G_{2} F_{+}
\end{pmatrix}
\begin{pmatrix}
     |\psi_{1} \rangle\\
    |\psi_{3} \rangle 
\end{pmatrix}
	=
	\begin{pmatrix}
     |\psi_{1} \rangle\\
    |\psi_{3} \rangle 
\end{pmatrix},
\end{equation}
and can be decoupled by repeating the scheme to find two extra reduced eigenvalue problems.

\section{Number of Parameters in a MPS representation of TC Model}\label{app_TCMPS}
\begin{figure}
  \centering
  \includegraphics[scale=0.3]{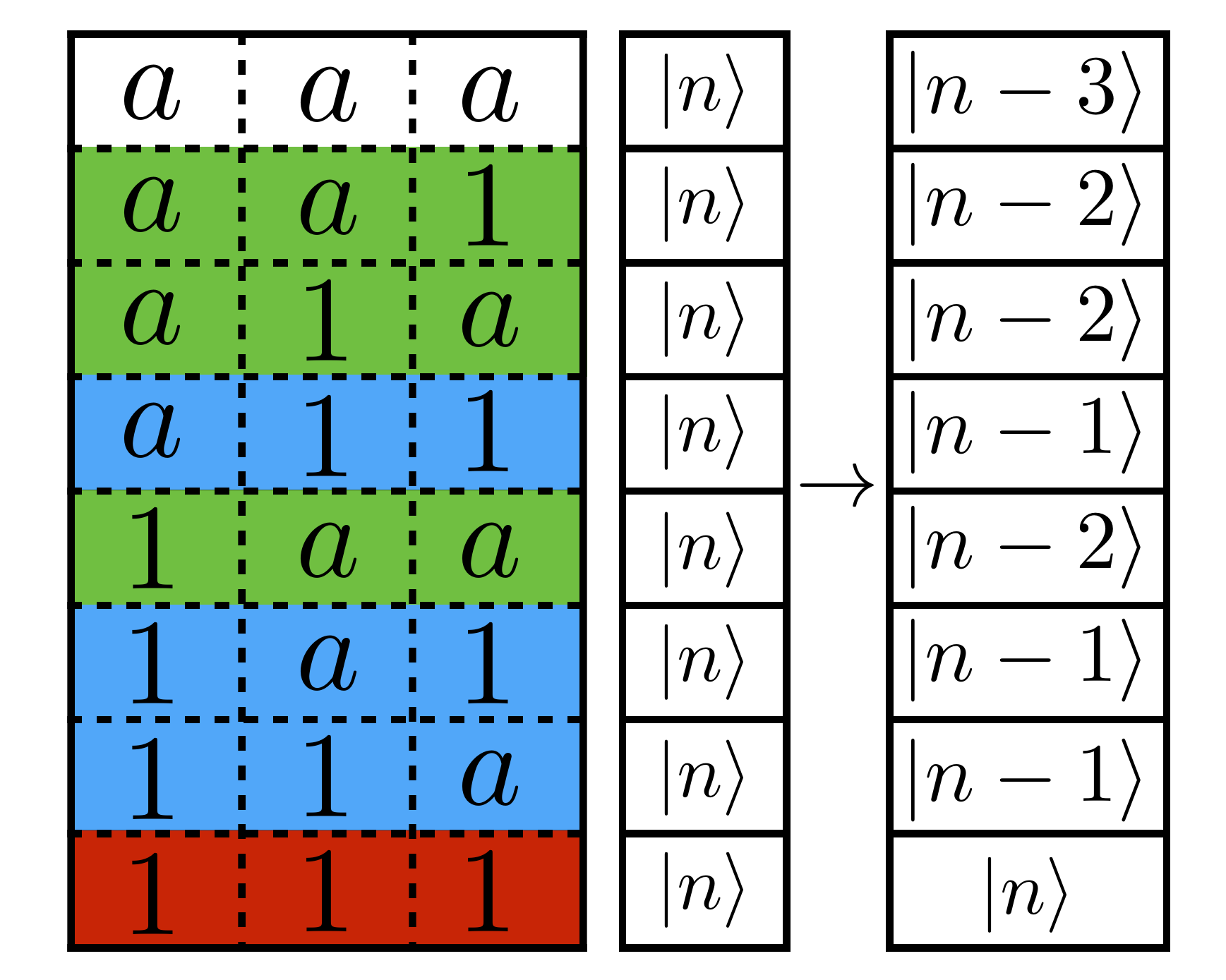}
  \caption{(color online) The vectorial representation of $\mathcal{M}^{\scriptscriptstyle  (N)}_{  n}$ in Eq.~\eqref{eq_TCMPS01} for $N=3$. The column vectors represent the diagonal terms of the matrices in $\mathcal{M}^{\scriptscriptstyle  (3)}_{  n}$. The rows in which the same number of $a$ appears are shown by the same color.}
  \label{figs_MPSTC}
\end{figure}
Let us first illustrate how the number of non-zero components of an eigenstate of TC model in the MPS representation can be counted for $N=3$, and then generalize the result for arbitrary $N$. $\mathcal{M}^{\scriptscriptstyle  (N)}_{  n}$ in Eq.~\eqref{eq_TCMPS01} is a multiplication of diagonal matrices whose non-zero elements can be depicted by column vectors as shown in Fig. \ref{figs_MPSTC}. The number of states $|n- j\rangle$ with $j=0,1,2,3$ in the ansatz is the binomial coefficient
$	 \begin{pmatrix}
      3   \\
        j
\end{pmatrix}$.
In fact, this number corresponds the number of ways to distribute $j$ operators in three consecutive entries of a row in Fig.~\ref{figs_MPSTC}. For instance, one (two) operator(s) can be distributed in three different ways which are shown in blue (green) in Fig.~\ref{figs_MPSTC}. The state $|n- j\rangle$ is vanished if $j>n$. If $n<3$, then
\begin{equation}
	P(3,n)=\sum_{j=0}^{n} \begin{pmatrix}
      3   \\
        j
\end{pmatrix},
\end{equation}
represents the number of all non-zero components. The same arguments applies for arbitrary $N$ which leads to the Eq.~\eqref{eq_dis}.

\section{Exact Diagonalization of $N$-spin Generalized Dicke models}\label{app_EDD}

Direct ED of the generalized Dicke model $H_{0}=a^{\dag} a$ and $F_{+}=\alpha a +\beta a^{\dag} +\gamma$
\begin{eqnarray}
H^{\left( N\right) }| \Psi ^{(N)} \rangle&=&\left(
\begin{array}{cc}
H_{+}^{\left( N-1\right) } & F_{+} \\
F_{-} & H_{-}^{\left( N-1\right) }%
\end{array}%
\right) \left(
\begin{array}{c}
|\psi _{+}^{\left( N-1\right) } \rangle \\
|\psi _{-}^{\left( N-1\right) } \rangle%
\end{array}%
\right) \nonumber \\
&=&\left(
\begin{array}{cc}
H^{\left( N-1\right) }+\Delta  & F_{+} \\
F_{-} & H^{\left( N-1\right) }-\Delta
\end{array}%
\right)
|\Psi^{(N)} \rangle
\nonumber
\\
&=& 
E |\Psi^{(N)} \rangle
\end{eqnarray} 
is about to diagonalize the matrix representation of $H^{\left( N\right) }$ in the truncated basis $\left\vert n,S_{i}\right\rangle $ where $n\in %
\left[ 0,M-1\right] $ is the boson occupation number and $S_{i}=\left( s_{1},s_{2},\ldots
s_{N}\right) _{i},i\in \left[ 1,2^{N}\right] $ is the collective spin index
of the N-spin system while $s_{i}$ is the state of a single spin. The matrix element of the Hamiltonian is simply given by%
\begin{equation*}
\left( H^{\left( N\right) }\right)^{m,n}_{i,j}=\left\langle
n,S_{i}\right\vert H^{\left( N\right) }\left\vert
m,S_{j}\right\rangle ,
\end{equation*}%
and the dimension of the matrix is $2^{N}\times M$.

\begin{figure}
  \centering
  \includegraphics[scale=0.7]{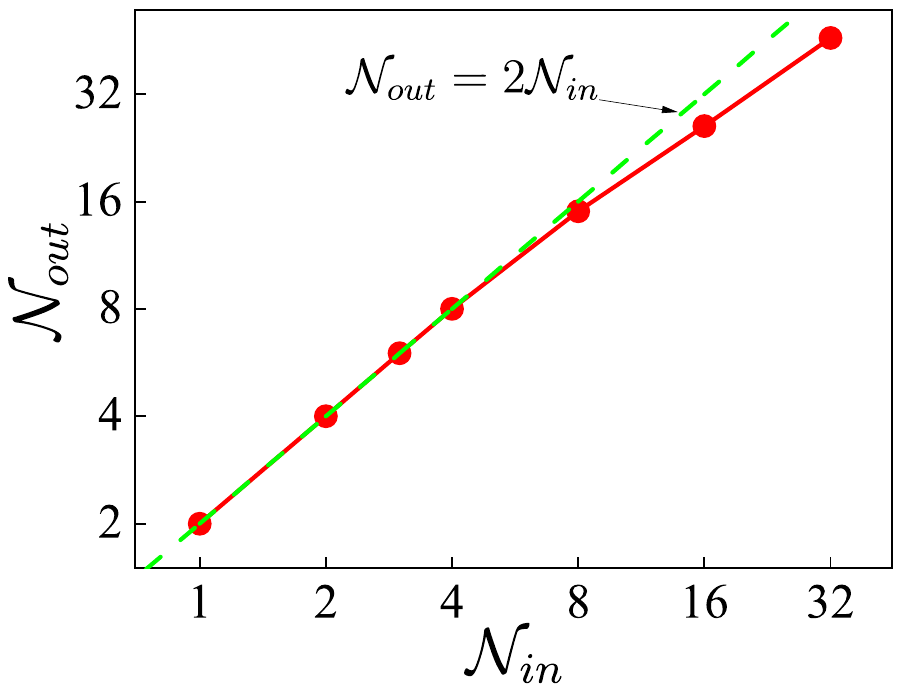}
  \caption{(color online) Results of generalized Dicke model of $N=7$ with $\alpha=0.15$, $\beta=0.1$, $\gamma=0.1$ and $\Delta=0.2$. The reduced Hamiltonian is diagonalized by means of $\mathcal{N}_{in}$ lowest eigenstates of $H_{-}^{\left( N-1\right) }$, and the number of extracted eigenstates of $H^{(N)}$, having energy errors 
  smaller than $4\%$ in comparison with a direct ED, is shown by $\mathcal{N}_{out}$. The dashed line corresponds to $\mathcal{N}_{out}/\mathcal{N}_{in}=2$ which is given as a matter of comparison.  }
  \label{figs1}
\end{figure}

Instead, the reduced Hamiltonian of the $N$-spin generalized Rabi model in the associated eigenproblem
\begin{eqnarray}
&&h_{+}^{(N-1)}|\psi _{+}^{\left( N-1\right) } \rangle= \nonumber \\
&&\left( H^{\left( N-1\right) }+\Delta +F_{+}\frac{1}{E+\Delta -H^{\left(
N-1\right) }}F_{-}\right) |\psi _{+}^{\left( N-1\right) } \rangle\nonumber \\
&&=E|\psi _{+}^{\left( N-1\right) } \rangle
\end{eqnarray}%
can be diagonalized by means of a truncated basis constructed out of eigenstates of $H_{-}^{\left( N-1\right) }$. The eigenenergies are then given by the roots of the determinant of the
matrix%
\begin{eqnarray}
\left(\mathcal{F}^{(N)}\right)_{j,k} 
=
\langle \Psi _{j}^{(N-1)} |(h_{+}^{(N-1)}-E) | \Psi
_{k}^{(N-1)}\rangle,
\label{eq_MEs}
\end{eqnarray}
where $\vert \Psi _{i}^{(N-1)}\rangle $ is the eigenstate of the 
$(N-1)$-spin Hamiltonian with its eigenenergy $E_{i}^{(N-1)}$. The eigenstate in a truncated basis is formally written as
\begin{eqnarray}
\left\vert \Psi _{i}^{N-1}\right\rangle 
 \approx \left(
\begin{array}{c}
 \sum_{n=0}^{M-1}c_{n}^{i,1}\left\vert
n\right\rangle  \\
\sum_{n=0}^{M-1}c_{n}^{i,2}\left\vert
n\right\rangle  \\
 \sum_{n=0}^{M-1}c_{n}^{i,3}\left\vert
n\right\rangle  \\
\ldots  \\
\sum_{n=0}^{M-1}c_{n}^{i,2^{N-1}}\left\vert n\right\rangle
\end{array}%
\right),
\end{eqnarray}%
where $c_{n}^{i,1}$ can be obtained by direct ED of $(N-1)$-spin system or obtained by solving its reduced eigenproblem. $h_{\pm }^{\left( N-1\right) }$ is represented in a matrix $^{(N)}$ by means of a $\mathcal{N}$-dimensional basis. If the truncation is the same for different spins, then $\max \left( \mathcal{N}\right) =2^{N-1}M$. In Fig. \ref{figs1}, it is shown that engaging $\mathcal{N}$ eigenstates 
of $(N-1)$-spin system (to diagonalize the reduced Hamiltonian) results in more than $\mathcal{N}$ eigenstates of $N$-spin system with energy errors $< 4\%$. Particularly, each eigenstate of $N-1$ spins leads to two eigenstates of $N$-spin system when $\mathcal{N} \leq 4 $.

In fact, we note that $\mathcal{N}$ eigenstates of the $(N-1)$-spin subsystem input in the $\mathcal{F}$ matrix lead to $2\mathcal{N}$ eigenstates 
of the $N$-spin system. The reason why $\mathcal{N}_{out} < 2\mathcal{N}_{in}$ with $\mathcal{N}_{in} > 4$ in Fig. \ref{figs1} is due to the fact that the eigenstates with error $> 4\%$ are not counted. 

We give a physical argument on the relation between input and output numbers of states in the diagonalization of the reduced Hamiltonian. Assume $\alpha=\beta=0$, then the matrix elements in Eq.~\eqref{eq_MEs} just have diagonal non-zero terms
\begin{equation}
		\left(\mathcal{F}\right)_{j,j}=\left( E_{j}^{(N-1)}+\Delta -E\right)
+ \frac{|\gamma|^{2}}{%
E+\Delta -E_{j}^{(N-1)}}  .
\end{equation} 
The eigenenergies are the roots of $|\mathcal{F}|=\prod_{j=1}^{\mathcal{N}} \left(\mathcal{F}\right)_{j,j}=0$ which has $2 \mathcal{N}$ solutions. Now, turning on $\alpha$ and $\beta$ adiabatically, the eigenenergies are redistributed and no state is generated/annihilated (as far as $H^{(N)}$ and $H^{(N-1)}- \Delta$ are not degenerate). Therefore, exploiting $ \mathcal{N}$ states to diagonalize $(N-1)$-spin system gives rise to $2 \mathcal{N}$ eigenstates of $N$-spin system.

\end{document}